\documentclass{aa}  

\usepackage{graphicx}
\usepackage{txfonts}
\usepackage{natbib,twoopt}
\usepackage[breaklinks=true]{hyperref} 
\hypersetup{
	bookmarks=false,         
	unicode=false,          
	pdftoolbar=true,        
	pdfmenubar=true,        
	pdffitwindow=false,     
	pdfstartview={FitH},    
	pdftitle={My title},    
	pdfauthor={Author},     
	pdfsubject={Subject},   
	pdfcreator={Creator},   
	pdfproducer={Producer}, 
	pdfkeywords={keyword1, key2, key3}, 
	pdfnewwindow=true,      
	colorlinks=true,       
	linkcolor=blue,          
	citecolor=blue,        
	filecolor=blue,      
	urlcolor=blue           
}

\bibpunct{(}{)}{;}{a}{}{,}             
\makeatletter
\newcommandtwoopt{\citeads}[3][][]{\href{http://adsabs.harvard.edu/abs/#3}%
	{\def\hyper@linkstart##1##2{}%
		\let\hyper@linkend\@empty\citealp[#1][#2]{#3}}}
\newcommandtwoopt{\citepads}[3][][]{\href{http://adsabs.harvard.edu/abs/#3}%
	{\def\hyper@linkstart##1##2{}%
		\let\hyper@linkend\@empty\citep[#1][#2]{#3}}}
\newcommandtwoopt{\citetads}[3][][]{\href{http://adsabs.harvard.edu/abs/#3}%
	{\def\hyper@linkstart##1##2{}%
		\let\hyper@linkend\@empty\citet[#1][#2]{#3}}}
\newcommandtwoopt{\citeyearads}[3][][]%
{\href{http://adsabs.harvard.edu/abs/#3}
	{\def\hyper@linkstart##1##2{}%
		\let\hyper@linkend\@empty\citeyear[#1][#2]{#3}}}
\makeatother

\begin{document} 

  \title{Radio continuum size evolution  of star-forming galaxies over $0.35<z<2.25$  }

   \author{E. F. ~Jim\'enez-Andrade 
          \inst{\ref{aifa}, \ref{imprs}} 
     	\and
     B.~Magnelli\inst{\ref{aifa}} \and A.~Karim\inst{\ref{aifa}} \and G.~Zamorani\inst{\ref{inaf}} \and M.~Bondi\inst{\ref{inaf_radio}} \and E.~Schinnerer\inst{\ref{hd}} \and M.~Sargent\inst{\ref{sussex}}  \and  E.~Romano-D\'iaz\inst{\ref{aifa}} \and M.~Novak\inst{\ref{hd}, \ref{zagreb}} \and P.~Lang\inst{\ref{hd}} \and F.~Bertoldi\inst{\ref{aifa}} \and E.~Vardoulaki\inst{\ref{aifa}}  \and S.~Toft\inst{\ref{dawn}} \and V.~Smol\v{c}i\'c\inst{\ref{zagreb}} \and K.~Harrington\inst{\ref{aifa}, \ref{imprs}} \and  S.~Leslie\inst{\ref{hd}}  \and J.~Delhaize\inst{\ref{zagreb}, \ref{capetown}}  \and D.~Liu\inst{\ref{hd}}  \and C.~Karoumpis\inst{\ref{aifa}, \ref{imprs}} \and J. Kartaltepe\inst{\ref{rochester}} \and A.M.~Koekemoer\inst{\ref{stsi}} }

   \institute{Argelander Institut f\"ur Astronomie, Universit\"at Bonn, Auf dem H\"ugel 71, D-53121 Bonn, Germany\label{aifa}\\
              \email{ericja@astro.uni-bonn.de}
         \and
             International Max Planck Research School of Astronomy and Astrophysics at the Universities of Bonn and Cologne, Bonn, Germany\label{imprs}
          \and  
          INAF – Osservatorio di Astrofisica e Scienza dello Spazio di Bologna, Via Gobetti 93/3, 40129 Bologna, Italy\label{inaf}
          \and             
           INAF – Istituto di Radioastronomia, Via Gobetti 101, 40129 Bologna, Italy\label{inaf_radio}
           \and
 		Max Planck Institute for Astronomy, K\"onigstuhl 17, 69117 Heidelberg, Germany\label{hd}        
 		\and 
 	Astronomy Centre, Department of Physics and Astronomy, University of Sussex, Brighton, BN1 9QH, UK\label{sussex}	
   		\and 
		 Department of Physics, University of Zagreb, Bijenička cesta 32, 10002 Zagreb, Croatia\label{zagreb} 
		\and 
		 Cosmic Dawn Center (DAWN),  Niels Bohr Institute, University of Copenhagen, Lyngbyvej 2, DK-2100 Denmark\label{dawn}
		 \and
		 Department of Astronomy, University of Cape Town, Private Bag X3, Rondebosch 7701, South Africa\label{capetown}
		\and
		School of Physics and Astronomy, Rochester Institute of Technology, 84 Lomb Memorial Drive, Rochester, NY 14623, USA\label{rochester}
		\and 
		Space Telescope Science Institute, 3700 San Martin Dr., Baltimore MD 21218, USA\label{stsi} }

   \date{Received, 2019; accepted 2019}

 
  \abstract
  	{To better constrain the physical mechanisms driving star formation, we present  the first systematic study of the radio continuum size evolution of star-forming galaxies (SFGs) over the redshift range $0.35<z<2.25$.  We use the VLA COSMOS 3GHz map (noise $\rm rms=2.3\,\mu Jy \,beam^{-1}$, $\theta_{\rm beam}=0.75\,\rm arcsec$) to construct  a mass-complete sample of 3184 radio-selected SFGs  that reside on and above the main-sequence (MS) of SFGs.  We constrain the overall extent of star formation activity in galaxies  by applying a 2D-Gaussian model to their radio continuum emission. Extensive Monte Carlo simulations are used to validate the robustness of our measurements and characterize the selection function. We find no clear dependence between the radio size and stellar mass, $M_{\star}$, of SFGs with $10.5\lesssim\log(M_\star/\rm M_\odot)\lesssim11.5$.  Our analysis suggests that MS galaxies are preferentially extended, while SFGs above the MS are always compact. The median effective radius of SFGs on (above) the MS of $R_{\rm eff}=1.5\pm0.2$ ($1.0\pm0.2$) kpc  remains nearly constant with cosmic time; a parametrization of the form $R_{\rm eff}\propto(1+z)^\alpha$ yields a shallow slope of only  $\alpha=-0.26\pm0.08\,(0.12\pm0.14)$ for SFGs on (above) the MS. The size of the stellar component of galaxies  is larger than  the extent of  the radio continuum emission by a  factor  $\sim$2 (1.3) at $z=0.5\,(2)$, indicating star formation is enhanced at small radii. The galactic-averaged star formation rate surface density $(\Sigma_{\rm SFR})$ scales with the distance to the MS,   
  	 except for a fraction of MS galaxies ($\lesssim10\%$) that harbor starburst-like $\Sigma_{\rm SFR}$. These ``hidden'' starbursts might have experienced a compaction phase due to disk instability and/or merger-driven burst of star formation, which may or may not significantly offset a galaxy from the MS.  We thus propose to jointly use $\Sigma_{\rm SFR}$ and distance to the MS to better identify the galaxy population undergoing a starbursting phase.   }


 	\titlerunning{Radio continuum size evolution of SFGs}
	\authorrunning{E.F. Jim\'enez-Andrade et al. }

   \maketitle
%

\section{Introduction}

Most galaxies follow a tight correlation in the star formation rate (SFR) -- stellar mass ($M_\star$) plane, which is known as the Main-Sequence (MS) of star-forming galaxies \citep[SFGs; e.g., ][]{brinchmann04, noeske07, elbaz07, salim07, daddi07, pannella09, magdis10, peng10, gonzalez10, rodighiero11, karim11, wuyts11, bouwens12, whitaker12, whitaker14,  rodighiero14,  pannella15, renzini15, schreiber15, schreiber17}.  This relation  holds over $\sim$ 90\% of the cosmic history of the Universe   \citep[e.g.,][]{stark13, gonzalez14,  steinhardt14, salmon15} and has a slope and normalization that increase with redshift, yet its dispersion of only $0.3$\,dex remains nearly constant throughout  cosmic time \citep[see][ and references therein]{speagle14, pearson18}.

Although most galaxies have an implied SFR that scatters within a factor two around the MS, some do show a significantly higher SFR. Those objects also exhibit a higher gas content, shorter gas depletion times  \citep[e.g., ][]{genzel15, schinnerer16, tacconi13, tacconi18} and  higher dust temperature \citep[e.g., ][]{magnelli14}.  Likewise, the stellar-light  radial distribution is different in these two galaxy populations; while  MS galaxies are well approximated by exponential disks \citep[e.g.,][]{bremer18}, those above (and below) it exhibit a higher central mass concentration  \citep[e.g.,][]{wuyts11}. Based on this dichotomy and the parametrization of the MS over cosmic time, a scenario has been proposed to explain the evolutionary path of galaxies along the MS. Since the normalization of the MS, gas fraction of galaxies  and cosmic molecular gas density  decrease  from $z\sim2.5$ to 0 at similar pace  \citep[e.g.,][]{speagle14, decarli16, tacconi18}, it is thought that
MS galaxies evolved through a steady mode of star formation, possibly regulated by the accretion of cool gas  from the intergalactic medium \citep[e.g.,][]{dekel09, keres09a, dave10, hodge12, romano-diaz14, romano-diaz17,  feng15, anglesalcazar17}. From theoretical predictions,  the scatter of the MS could hence be explained as the result of a fluctuating gas inflow rate, that is  different in each galaxy \citep[e.g., ][]{tacchella16, mitra17}. In this context, a galaxy  enhances its SFR and moves towards the upper envelope  of the MS due to gas compaction. As the gas is depleted, the SFR decreases and the galaxy falls below the MS.  
As long as  a SFG is replenished with fresh gas, within a timescale shorter than its depletion time, it will be confined within the MS \citep{tacchella16}. On the other hand, the enhanced star formation efficiency of galaxies above  the MS has been linked to  mergers  \citep[e.g., ][]{walter09, narayanan10, hayward11, alaghband12,riechers13, riechers14} and instability episodes in gas-rich disks  \citep[particularly at high redshift; e.g., ][]{dave10, hodge12, wang18}. 

 A crucial parameter to verify these scenarios is the size of a galaxy. 
 Recent studies have explored the structural properties of SFGs by mapping their stellar component \citep[e.g., ][]{vanderwel14, shibuya15, mowla18}.  However, the size of the overall star-forming component has been poorly explored.  This is partially due to observationally expensive high-resolution IR/radio observations, which have been limited to relatively small  samples of SFGs \citep[e.g., ][]{rujopakarn16, miettinen17, murphy17, elbaz18}. While large and representative samples of SFGs can be obtained from UV/optical observations, these are affected by dust extinction, rendering size measurement difficult \citep[e.g., ][]{elbaz11, nelson16}.  
 To better understand the  mechanisms that regulate star formation in galaxies  we need a statistically significant, mass-complete sample of radio-selected SFGs over cosmic time, and  a dust-unbiased  measure of the size of the star formation activity in galaxies.

The cm wavelength radio emission has been established as a proxy of the massive SFR in galaxies, both locally and at high redshift  \citep[e.g., ][]{bell03, garn09}. Empirically, this is evidenced by a strong correlation between the radio flux density and the far-infrared (FIR) flux \citep[e.g., ][]{helou85, yun01, murphy06, murphy06b, murphy09,  murphy12, sargent10, magnelli15, delhaize17}. This can be understood in that the stellar UV radiation is mostly absorbed by dust that re-emits this energy in the FIR. On the other hand, supernova explosions of the same massive stars give rise to relativistic electrons emitting radio synchrotron radiation \citep{helou93}. This radio emission is not affected by extinction, and with radio interferometers it can be imaged over wide fields at a resolution much better than is currently possible in the FIR or sub-mm.

The latter has motivated the   VLA COSMOS 3GHz imaging survey \citep{smolcic17} that reached an unprecedented resolution and sensitivity ($\theta_{\rm beam}=0.75\,\rm arcsec$, noise $\rm rms=2.3\,\mu Jy \, beam^{-1}$) over the two square degrees of the COSMOS field, enabling size measurements for a large number of radio sources in the $\mu \rm Jy$ regime \citep{bondi18}. Over the redshift range explored here, this survey allows us to sample  the  rest-frame frequency range $4\lesssim \nu \lesssim 10\,\mathrm{GHz}$, which is dominated by the steep-spectrum of synchrotron radiation of SFGs \citep[e.g., ][]{murphy09}. 
In combination with  reliable photometric redshifts and stellar mass 
content measurements accumulated in the COSMOS 2015 catalog \citep{laigle16},  we are  able to study the radio size evolution over $0.35<z<2.25$ of a mass-complete sample of radio-selected SFGs. 

In this work,  we investigate how the radio continuum size of a SFG relates to its stellar mass, size of its stellar component and distance to the MS:
\begin{equation}
 \Delta\log({\rm SSFR})_{\rm MS}=\log[{\rm SSFR}_{\rm galaxy}/{\rm SSFR}_{\rm MS}(M_\star,z)]
\end{equation}
\noindent
 where ${\rm SSFR=SFR}/M_{\star}$ is the specific SFR of a galaxy. In particular, by exploring the relation between  the galactic-averaged star formation surface density ($\Sigma_{\rm SFR}$) and  $\Delta\log({\rm SSFR})_{\rm MS}$, we aim to verify if galaxies harboring intense star formation activity experience a compaction phase --  as predicted by cosmological simulations \citep[e.g., ][]{tacchella16} and observed in small samples of SFGs \citep[e.g., ][]{rujopakarn16, elbaz18}.  

This paper is structured as follows. In Sect. \ref{sec:data} we present the VLA COSMOS 3GHz map and the COSMOS2015 catalog, both used to identify the SFGs studied in this work.  The sample selection and the methodology to test the robustness of our measurements are given  in Sect. \ref{sec:dataanalysis}. In Sect. \ref{sec:results}, we present  the radio size$-$stellar mass, radio size$-\Delta\log(\rm SSFR)_{\rm MS}$  and $\Sigma_{\rm SFR}-\Delta\log(\rm SSFR)_{\rm MS}$ relations; as well as the redshift evolution of the radio continuum size of  SFGs with $10.5\lesssim\log(M_\star/\rm M_\odot)\lesssim11.5$. Results are discussed in Sect. \ref{sec:discussion}, while a summary is given in Sect. \ref{sec:summary}. Throughout, we assume a cosmology of $h_0 = 0.7$, $\Omega_M = 0.3$, and $\Omega_\Lambda = 0.7$.\\

\section{Data} \label{sec:data}

\subsection{VLA COSMOS 3GHz Large Project}\label{subsec:vlaproject}

The  VLA COSMOS 3 GHz survey \citep{smolcic17} consists of 384 hr of observations  (A array --  324 h,  C array  -- 60h) with the {\it Karl G. Jansky} Very Large Array. A total of 192 individual pointings (HPBW=15\,arcmin) were performed to achieve a  uniform rms over the two square degrees COSMOS field. Data calibration was performed with AIPSLite \citep{bourke14}. The imaging was done via the  multi-scale multifrequency synthesis (MSMF) algorithm in CASA,  using a robust parameter of 0.5 to obtain the best possible combination between resolution and sensitivity. Given the large data volume of the observations, joint deconvolution of the 192 pointings was unpractical.  Therefore, each pointing was imaged individually using a circular restored beam with a FWHM of 0.75 arcsec. The final mosaic was produced using a noise weighted mean of all the individually imaged pointings, reaching a median  rms of 2.3\,$\rm \mu Jy\, \rm{beam}  ^{-1}$.

\subsection{COSMOS2015 catalog}\label{subsec:cosmos2015catalog}

The COSMOS2015 catalog comprises  photometric redshifts and stellar masses for more than half a million galaxies  over the two square degrees of the COSMOS field \citep{laigle16}. This  near-IR-selected catalog  combines  extensive deep photometric information from the YJHKs images of the UltraVISTA \citep{mccracken12} DR2  survey, Y-band images from Subaru/Hyper-Suprime-Cam \citep{miyazaki12}, and infrared data from the Spitzer Large Area Survey (SPLASH) within the Hyper-Suprime-Cam Spitzer legacy program.

Photometric redshifts were derived with {\sc LePhare} \citep{arnouts02, ilbert06} using a set of 31 templates of  spiral and elliptical galaxies from \cite{polletta07}, as well as   12 templates of young blue SFGs using the  \cite{bruzual03} models.  
Through a comparison with  spectroscopic redshift samples in the COSMOS field, \cite{laigle16} derived a  photometric redshift precision  of $\sigma_{\Delta_z}/(1+z_{s})=0.007$  and a low catastrophic failure fraction of $\eta= 0.5\%$ for $z_s<3$.  

Stellar masses were also derived  with  {\sc LePhare} using a library of synthetic spectra from the Stellar Population Synthesis model of \cite{bruzual03}. A \cite{chabrier03} initial mass
function, an exponentially declining and delayed SFH and  solar/half-solar metallicities were considered. The stellar masses used here corresponds to the median of the inferred probability distribution function.  A 90\% completeness limit of $10^{8.5}\,(10^{10})\, \rm M_\odot$ was achieved up to $z = 0.35 \,(2.25)$.

\section{Data analysis}\label{sec:dataanalysis}

We  measure the size and flux density of radio sources directly from the VLA COSMOS 3GHz mosaic, i.e.  in the image plane, and further revise those estimates using extensive Monte Carlo simulations. While these sizes and fluxes could also be estimated in the uv-plane, this is impractical due to the large data volume of the VLA COSMOS 3 GHz survey.

\subsection{Source extraction}\label{subsec:sourceextraction}

The advent of large radio  astronomical surveys  has stimulated the  development of robust  source extraction algorithms such as {\tt blobcat} \citep{hales12} and {\tt PyBDSF} \citep{mohan15}.  Here, we use {\tt PyBDSF}  as it provides  parametric  information of the source morphology such as the deconvolved major axis {FWHM} ($\theta_{\rm M}$),  that is, 
\begin{equation}
\theta_{\rm M}=\left( \left(\theta_{\rm M}^{\rm obs}\right)^2-\left(\theta_{\rm beam}\right)^2 \right)^{\frac{1}{2}},
\end{equation}
\noindent
where $\theta_{\rm beam}$ is the {FWHM} of the synthesized beam of the VLA COSMOS 3GHz map (0.75 arcsec) and $\theta_{\rm M}^{\rm obs}$ the observed/convolved major axis {FWHM}.

 {\tt PyBDSF} characterizes the radio source properties as follows. First, it identifies peaks of emission above a given threshold  ({\tt thresh$\_$pix}) that are surrounded by contiguous pixels, i.e.  islands, with emission greater than a minimum value  ({\tt thresh$\_$isl}).  Second, it fits  multiple Gaussians to each island depending on the number of the peaks identified within it. Finally, Gaussians are grouped into sources if (a) their centers are separated by a distance less than half of the sum of their FWHMs and (b) all the pixels on the line joining their centers have a value greater than {\tt thresh$\_$isl}. The total flux of the sources is estimated by adding those from the individual Gaussians, while the central position and source size are determined via moment analysis. The error of each fitted parameter is computed using the formulae in \cite{condon97}.

We run {\tt PyBDSF} over the VLA COSMOS 3 GHz  mosaic adopting  {\tt thresh$\_$pix}=5$\sigma$,  {\tt thresh$\_$isl}=3$\sigma$, and a minimum number of pixels in an island ({\tt minpix$\_$isl}) of 9 (as in \citet{smolcic17}). By selecting sources within the inner two square degrees of the COSMOS field, where the rms remains homogeneous,  we find 10078 sources.  Within the same area, there are 10689 sources in the catalog presented by \cite{smolcic17}, of which  9223 are also retrieved by {\tt PyBDSF}.  In the subsequent analysis, we use these matched sources to enhance the pureness of our radio source catalog.

 \subsection{AGN rejection}\label{subsec:agnrejection}
To identify  galaxies in which the radio continuum emission is associated with an active galactic nucleus (AGN) -- and not star formation -- we rely on the results from  \cite{smolcic17b}. They characterized the host galaxy of radio sources in the  VLA COSMOS 3 GHz  map by  identifying their optical/NIR/MIR counterparts from: the  i-band selected catalog \citep[optical;][]{capak07}, the COSMOS2015 catalog \citep[NIR; ][]{laigle16},  and the Spitzer COSMOS (S-COSMOS) Infrared Array Camera \citep[3.6 $\mu$m-selected, IRAC; ][]{sanders07}. Based on this multi-wavelength counterpart association, a sample of AGN and SFGs was assembled.

 AGN host galaxies were identified as such,  and excluded from our sample, if:

 \begin{itemize}
 	\item the intrinsic [0.5–8] keV  X-ray luminosity is greater than $L_X = 10^{42} \rm erg \rm{s}^{-1}$ \citep[e.g., ][]{szokoly04}, 
 	
 	\item  the flux throughout the four IRAC bands (3.6, 4.5, 5.8, and 8) displays a monotonic rise and follows the criterion proposed by \cite{donley12}, 
 	
 	 \item an AGN component significantly improves the  optical to millimetre SED fitting  \citep[as in ][]{dacunha08,berta13,delvecchio14,delvecchio17},
 	
 	\item $M_{NUV}-M_r$, i.e. rest-frame near ultraviolet (NUV) minus $r+$ band, is greater than 3.5  \citep{ilbert10}, 
 	
 	\item the observed radio emission $ L_{1.4\rm GHz}$ exceed that expected from the host galaxy $\rm SFR_{IR}$ \cite[estimated via IR SED fitting,][]{delvecchio17}.  
 \end{itemize}

 Excluding AGN hosts  through all  these criteria yields a highly clean sample of SFGs.  Within the redshift range probed in this work ($0.35<z<2.25$), we find that 4216 galaxies match with our catalog of 9223 radio sources and have available stellar mass estimates in the COSMOS2015 catalog. While most of them (3248, i.e. 77\%) are classified as SFGs,  968 galaxies (23\%) exhibit one or more of the aforementioned signatures of  AGN activity.  Since comparing the radio size evolution of AGN and SFGs is beyond the scope of this work, we refer the reader to \citet{bondi18} who presented such an analysis using the VLA COSMOS 3GHz map -- following the same   AGN-SFGs classification scheme  used here.

We note that out of the 3248 radio-selected SFGs, 64 (2\%) of them  are fitted with multiple Gaussians by {\tt PyBDSF}, suggesting a more complex and/or extended morphology.  Since  modeling such systems in our Monte Carlo simulations  (Sect. \ref{subsec:mcsimulations}) is challenging,  we  exclude them from the analysis. We verified, however, that none of the relations/results reported thereafter are  affected, within uncertainty, by the inclusion of these multi-component sources. Our final sample, therefore, comprises 3184 radio-selected  SFGs  over the redshift range $0.35<z<2.25$, in which a mass-complete sample of $\log(M_\star/M_\odot)\gtrsim10.5$ SFGs can be assembled (Sect. \ref{subsec:finalsample}).

\subsection{Accuracies and limitations of our size and flux density measurements}\label{subsec:mcsimulations}

In this section, we describe the Monte Carlo (MC) simulations used to   characterize the biases associated with size and flux determination of SFGs in the sample.   This approach is based on the injection of mock sources, following a realistic flux and size distribution, into noise maps that accurately represent the original dataset \citep[e.g., ][Sect. 3.2]{casey14}.   After retrieving these sources from the maps with {\tt PyBDSF}, we compare the input and output properties and hence address these particular questions: (a) what are the minimum/maximum source sizes we can detect in the VLA COSMOS 3GHz mosaic at a given flux density? and (b) how reliable are our measurements for a given intrinsic flux density  and FWHM?

\begin{figure}
	\centering
	\includegraphics[width=8.cm]{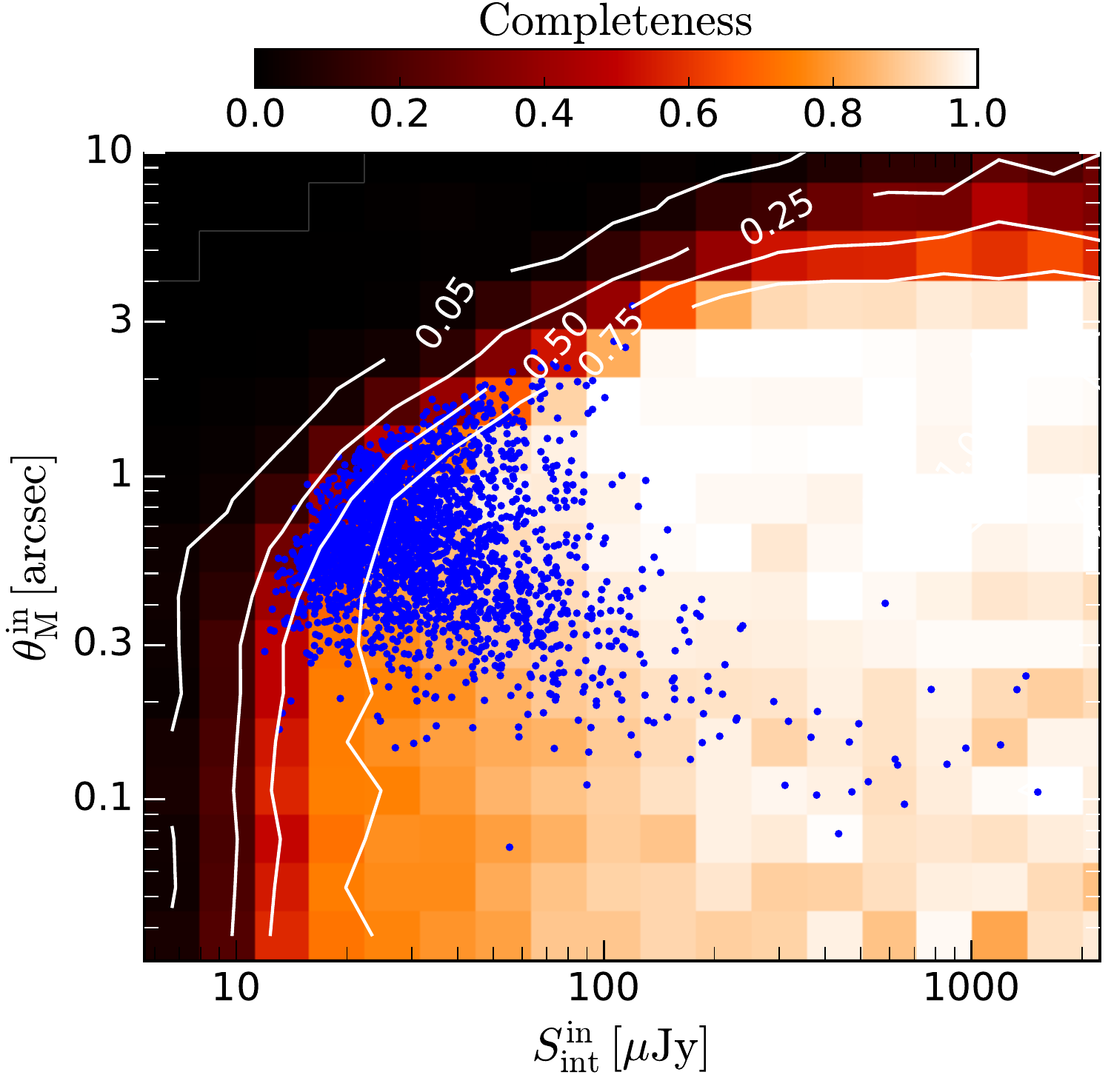}
	\caption{Completeness in the $\theta_{\rm M}^{\rm in}$ vs $S_{\rm int}^{\rm in}$ plane as inferred from extensive MC simulations.
		The completeness  given by the color scale represents the fraction of sources recovered by \texttt{PyBDSF} (resolved and unresolved) 	over the original number of mock sources. The blue points show  the position of  resolved SFGs in the VLA COSMOS 3GHz map studied in this work. White contours represent a completeness levels of 5, 25,  50 and 75\%.  }  
	\label{fig:completeness}%
\end{figure}

These MC simulations  require a mock sample that follows the intrinsic, yet unknown, flux density $(S_{\rm int})$ and angular size $(\theta_{\rm M})$ distributions of SFGs.  For this purpose, we use previous constraints on the $\mu\rm Jy$ radio source population as presented in \cite{smolcic17}.  First, we  approximate  the observed  flux density distribution  of this mock sample with  a single power-law model ($N\propto S_{\rm int}^{-0.8}$). Second, we assume that their angular size is linked to their total flux density as \citep{windhorst90, richards00}:  $\theta_{\rm median}\,[{\rm arcsec}]=1.8  S_{\rm int}^{0.6} \, [\rm mJy]$  \citep{bondi03, smolcic17}.

The input sample comprises \hbox{$\sim7\times10^5$}  sources modeled with a single Gaussian component. We explore the parameter space where  $\theta_{\rm M}^{\rm in}$ ranges from  $0.03-12$ arcsec (with ellipticity $e=0.25,0.5,0.75,1$ and random position angle)   and $10^{-5}\,\rm Jy <S_{\rm int}\,\rm <10^{-1.5}$\,Jy; i.e. the observed range of  retrieved VLA-COSMOS 3GHz galaxies (see Fig. \ref{fig:completeness}). 
These mock galaxies were  convolved with the synthesized beam and  randomly  injected into the mosaic in purely noise dominated regions, i.e. those  areas where no original source is found within $36\times36\rm \, arcsec^2$.  They were subsequently retrieved with {\tt PyBDSF} using the same parameters described in Sect. \ref{subsec:sourceextraction} and  cross-matched with the input mock catalog (within a circle of  1 arcsec radius). The ratio of the number of successfully retrieved  mock sources over  original mock sources injected in the map, in each  $[S_{\rm int}^{\rm in},\, \theta_{\rm M}^{\rm in}]$ bin, represents the completeness (see Fig. \ref{fig:completeness})

\subsubsection{Selection function,  maximum recovered size}\label{subsubsec:selectionfunction}

To constrain the maximum  detectable size of a galaxy  as a function of  redshift, stellar mass and $\Delta\log(\rm SSFR)_{\rm MS}$, we explore the  angular size of mock sources that were resolved by {\tt PyBDSF}.   The completeness levels  in the $\theta_{\rm M}^{\rm in}$ vs  $S_{\rm int}^{\rm in}$ plane (Fig.  \ref{fig:completeness}) reveal that the maximum recovered deconvolved  { FWHM} ($\theta_{\rm M}^{\rm max}$), for sources within $10^{-5}\, \textrm{Jy}<S_{\textrm{int}}^{\textrm{in}}<10^{-3}\,\textrm{Jy}$, strongly depends on $S_{\rm int}^{\rm in}$; i.e.  a higher flux density  increases the possibility of detecting extended  sources\footnote{Not all  bins at the bright/compact-end exhibit a 100\% completeness. We attribute this result to the minimum number of pixels in an island ({\tt minpix$\_$isl=9}) used  to retrieve the radio sources with {\tt PyBDSF}. Negative noise fluctuations might hinder the detection of islands of emission above this threshold.  Certainly, we verified that using {\tt minpix$\_$isl}=6 yield a higher completeness at the bright/compact-end. Yet, we adopted {\tt minpix$\_$isl}=9 to be consistent with the original VLA COSMOS 3GHz catalog \citep{smolcic17}.}.  Thus, at a given redshift,   faint galaxies are preferentially detected if they are compact, while bright starbursting systems  are detected even if they are extended. This selection function (i.e. completeness level of 10\%)   is further discussed in  Sects. \ref{subsec:size-mass} \& \ref{subsec:size-dms}.

\subsubsection{Upper limit for the size of unresolved sources}\label{subsubsec:upperlimitunresolved}

\begin{figure}
	\centering
	\includegraphics[width=9.05cm]{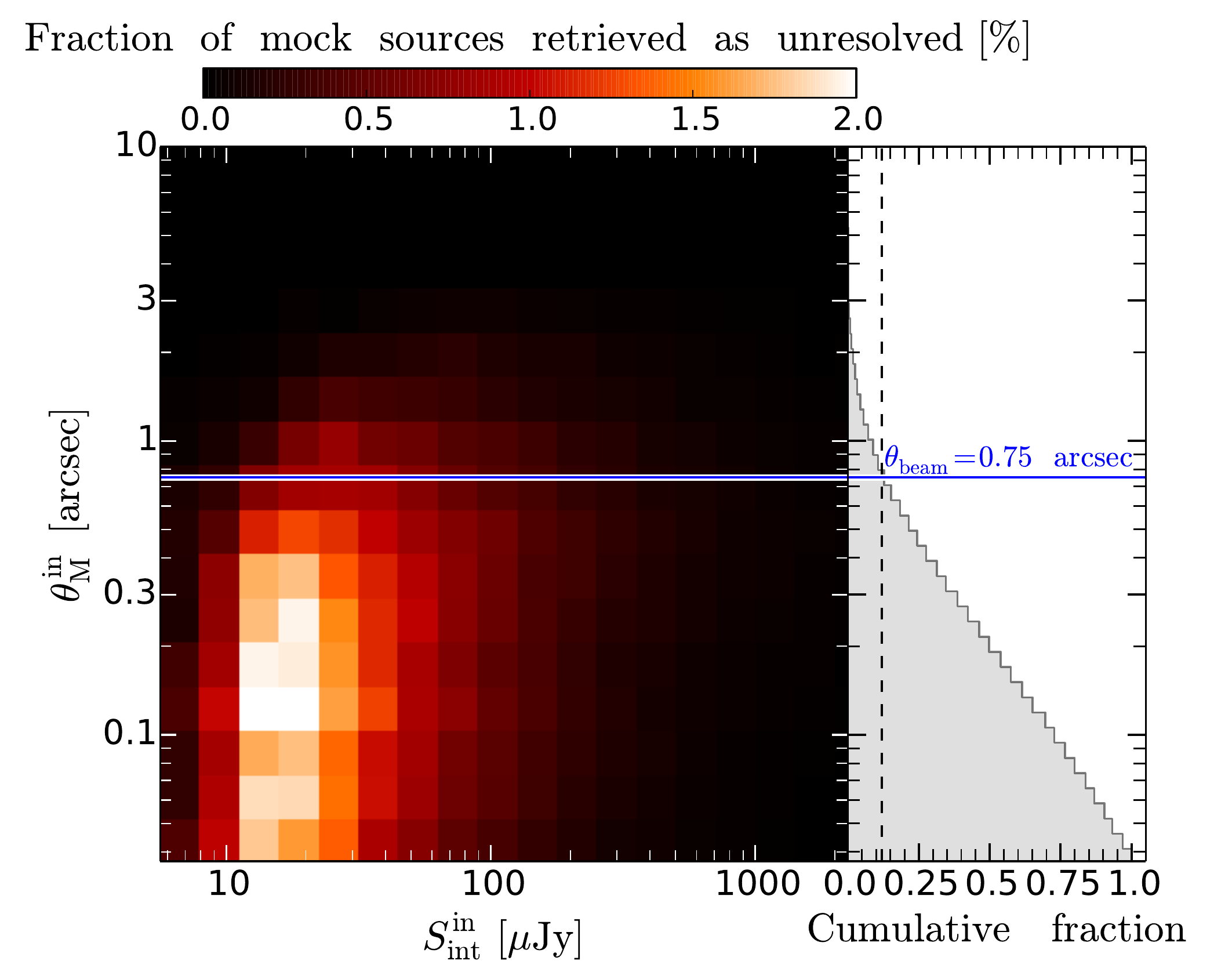}
	
	\caption{  {\it Left panel:} fraction of  mock sources retrieved by {\tt PyBDSF} as unresolved in the $\theta_{\rm M}^{\rm in}-S_{\rm int}^{\rm in}$ plane.   It shows the ratio between the number of unresolved sources per bin over the total number of  unresolved sources in the entire parameter space. 
		{\it Right panel:} cumulative size distribution of mock sources retrieved as unresolved, which represent 29\% of the total number of sources injected in our MC simulations. 
		Around $90$\% of them lie below $\theta_{\rm beam}=0.75\, \rm arcsec$ (blue line), hence we use this value as the upper limit for the size of unresolved SFGs  in the VLA COSMOS 3GHz map. 
	}
	\label{fig:upperlimitunresolved}%
\end{figure}

665 SFGs (21\%)  from our  sample  are unresolved $(\theta_{\rm M}^{\rm out} = 0 \rm \,  arcsec)$ by {\tt PyBDSF}.  In order to assign an upper limit to their intrinsic angular size ($\theta_{\rm M}^{\rm in} < \theta_{\rm lim}$), we explore the input  size of mock sources  retrieved as unresolved in the MC simulations. In Fig. \ref{fig:upperlimitunresolved}, we plot their distribution in the $\theta_{\rm M}^{\rm in}-S_{\rm int}^{\rm in}$ plane. Most of the sources retrieved as unresolved by {\tt PyBDSF} are, as expected, at the faint and compact-end of the parameter space tested here. 
Based on their angular size distribution (Fig. \ref{fig:upperlimitunresolved}, right panel),  we find that around 90\% of them satisfy the condition: $\theta_{\rm M}^{\rm in} \leq \theta_{\rm beam}$ (blue line).  We thus define  $\theta_{\rm beam}=0.75\, \rm arcsec$ as the upper limit for the size of the 665 unresolved SFGs in our sample.

\subsubsection{How reliable are the retrieved { FWHM} and flux density?}\label{subsubsec:corrections}

It is well-known that noise fluctuations ``boost" the flux of faint sources detected in sensitivity-limited astronomical survey
 \citep[e.g., ][]{hogg98, coppin05,  casey14}. It is expected that a similar effect takes place when determining the size of faint and compact  sources. Therefore, in a pioneering effort, we use the MC simulations to correct both the {FWHM} and flux density (and associated uncertainties) provided by {\tt PyBDSF}. We proceed as follows:

\begin{figure*}
	\centering
	\includegraphics[width=18.5cm]{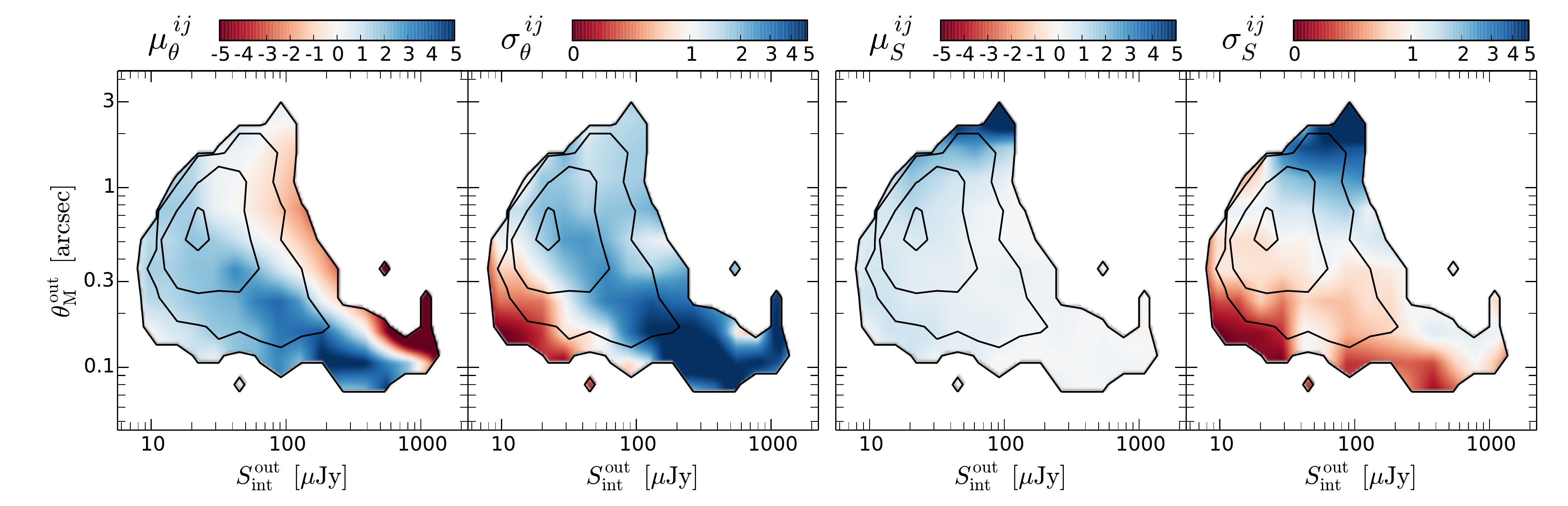}
	
	\caption{Systematic errors and uncertainties for the {FWHM} (left two panels) and flux density (right  two panels) of mock sources in the $\theta_{\rm M}^{\rm out}-S_{\rm int}^{\rm out}$ plane covered by 2519 resolved SFGs studied in this work. Contour levels showing the distribution of these sources are at 1, 5, 50, 250 sources per bin.   
		Positive/negative values of $\mu^{ij}_{\theta}$ and $\mu^{ij}_{S}$ indicate that the measured quantity is over/underestimated. Values of  $\sigma^{ij}_{\theta}$ and $\sigma^{ij}_{S}$  greater/lower than 1 suggest that the uncertainty of the measured parameter is being under/overestimated. 
} 
	\label{fig:sigma-mudistributions}%
\end{figure*}

\begin{enumerate}
\item We create a catalog containing all mock sources retrieved by {\tt PyBDSF}. Hence,  it contains information about the input ($S_{\rm int}^{\rm in}$, $\theta_{\rm M}^{\rm in}$) and output parameters ($S_{\rm int}^{\rm out}$, $\theta_{\rm M}^{\rm out}$). \\

\item All the sources in the catalog are binned in the $S_{\rm int}^{\rm out} - \theta_{\rm M}^{\rm out}$ plane (as shown in Fig. \ref{fig:completeness}). For all objects in each bin, we estimate $r_{\theta}\equiv(\theta_{\rm M}^{\rm out}-\theta_{\rm M}^{\rm in})/\sigma_{\theta}$ and/or $r_{S}\equiv(S_{\rm M}^{\rm out}-S_{\rm M}^{\rm in})/\sigma_{S}$, where $\sigma_{\theta}$ and $\sigma_{S}$ are the uncertainties provided by {\tt PyBDSF}. \\

\item We derive the mean $(\mu)$ and standard deviation $(\sigma)$ of the $r_{\theta}$ and  $r_{S}$  distributions  (Fig. \ref{fig:sigma-mudistributions}).  While the value of $\mu$  quantifies systematic biases (e.g., ``flux boosting''),  $\sigma$ evaluates whether the uncertainties  given by {\tt PyBDSF} are under/overestimated. \\

In the ideal case where the measured properties and uncertainties are an appropriate description of the input mock sources, the mean $(\mu)$ and standard deviation $(\sigma)$ of the distribution should be 0 and 1, respectively. Nevertheless, for both { FWHM} and flux density, $\mu$ is generally larger than zero (Fig. \ref{fig:sigma-mudistributions}), meaning that {\tt PyBDSF} tends to overestimate the size and flux density of mock sources.  The value of $\sigma$ is also heterogeneous across the $\theta_{\rm M}^{\rm out} - S_{\rm int}^{\rm out}$ plane (Fig. \ref{fig:sigma-mudistributions});   $\sigma^{ij}_{\theta}$ or $\sigma^{ij}_{S}$  greater/lower than 1 suggest that the uncertainty provided by {\tt PyPDSF} is being under/overestimated. \\

\item  Under the condition that all $r_\theta$ and $r_S$ distributions should have a mean of zero and dispersion of 1, the  corrected source properties $( \theta_{\rm M}^{\rm ' out},\, S_{\rm int}^{\rm ' out} )$ and associated uncertainties $(\sigma_{\theta}',\, \sigma_{S}')$ are given by:
\begin{equation}
\left.
\begin{aligned}
\theta_{\rm M}^{\rm ' out}=\theta_{\rm M}^{\rm out} - \mu^{ij}_{\theta}\times\sigma_{\theta}, \hspace{0.1cm} \\ 
\hspace{2.5cm} S_{\rm int}^{\rm ' out}=S_{\rm int}^{\rm out} - \mu^{ij}_{S}\times\sigma_{S},
\end{aligned}
\hspace{0.5cm}\right\}
\end{equation}
\noindent
and 
\begin{equation}
\left.
\begin{aligned}
\sigma_{\theta}'=\sigma_{\theta}^{ij}\times \sigma_{\theta},  \\ 
\hspace{3.5cm} \sigma_{S}'=\sigma_{S}^{ij}\times \sigma_{S},
\end{aligned}
\hspace{0.5cm}\right\}
\end{equation}
\noindent
where  $\mu^{ij}_{\theta}, \, \mu^{ij}_{S}, \sigma^{ij}_{\theta}$ and $\sigma^{ij}_{S}$ are the mean and standard deviations of the  $r_\theta$ and $r_S$  distributions  in each bin; where $i=1...m$ and $j=1...n$, with $m$ and $n$ the numbers of columns and rows used to grid the $\theta_{\rm M}^{\rm out} - S_{\rm int}^{\rm out}$ plane. \\

\begin{figure}
	\centering
	\includegraphics[width=9.cm]{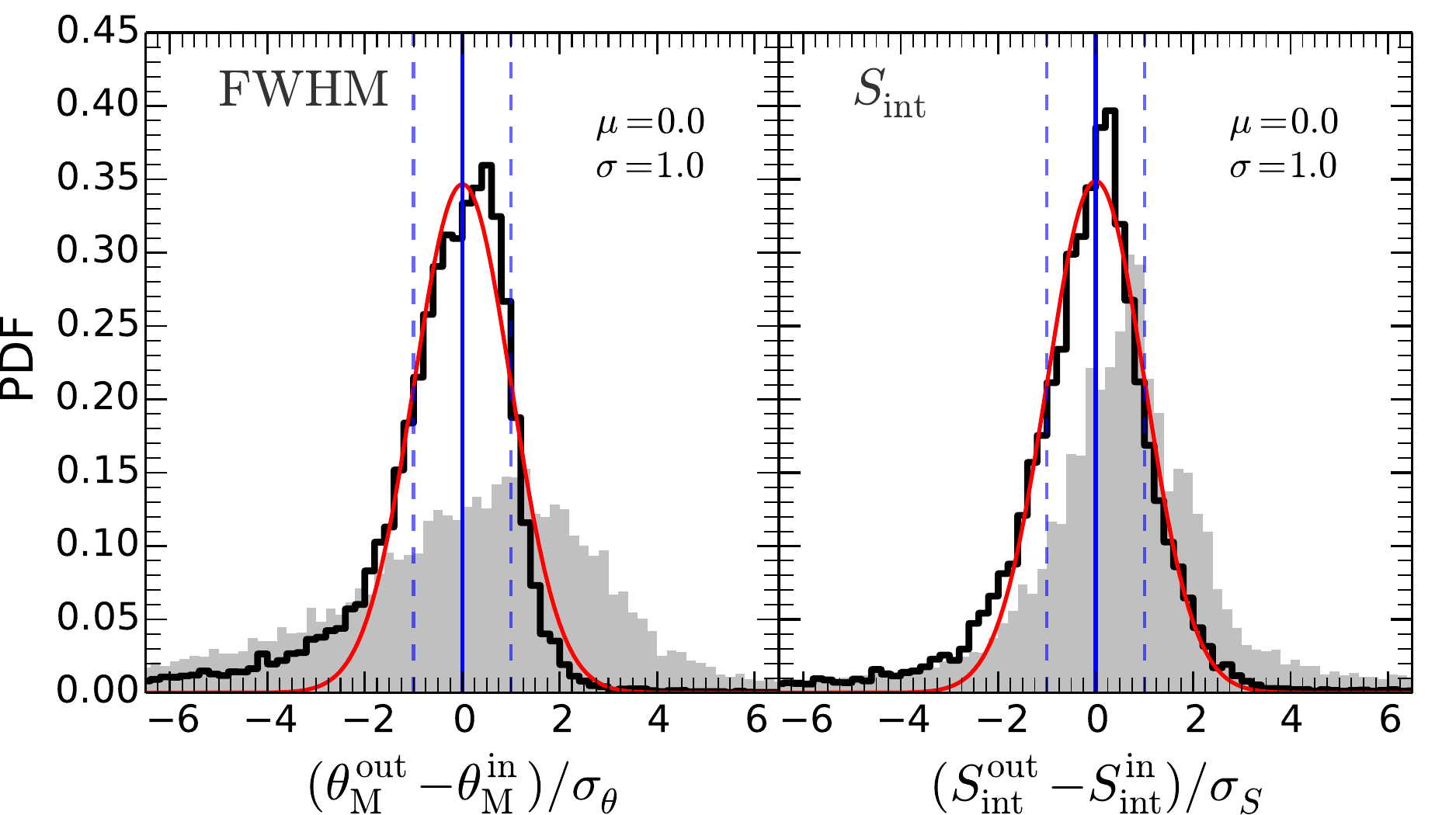}
	\caption{Distribution of sigma deviations for the FWHM (left panel) and  integrated flux density (right panel) of all mock sources. The distribution that is produced from the corrected quantities is shown in black, while  in gray the one  that obtained from the measured/original quantities given by {\tt PyBDSF}. A single component Gaussian fit is shown in red. For both corrected distributions, we find the best fitting parameters  of  $\mu=0$ and $\sigma=1$ (blue solid and dashed lines), which indicates that the corrected flux density  and {FWHM} (and associated uncertainties) are a proper description of the mock sources.  Blue solid (dashed) lines illustrate the locus of  $\mu=0$ ($\sigma=1$).    	
	}
	\label{fig:sigmadistributions}%
\end{figure}
\begin{figure*}
	\centering
	\includegraphics[width=18.5cm]{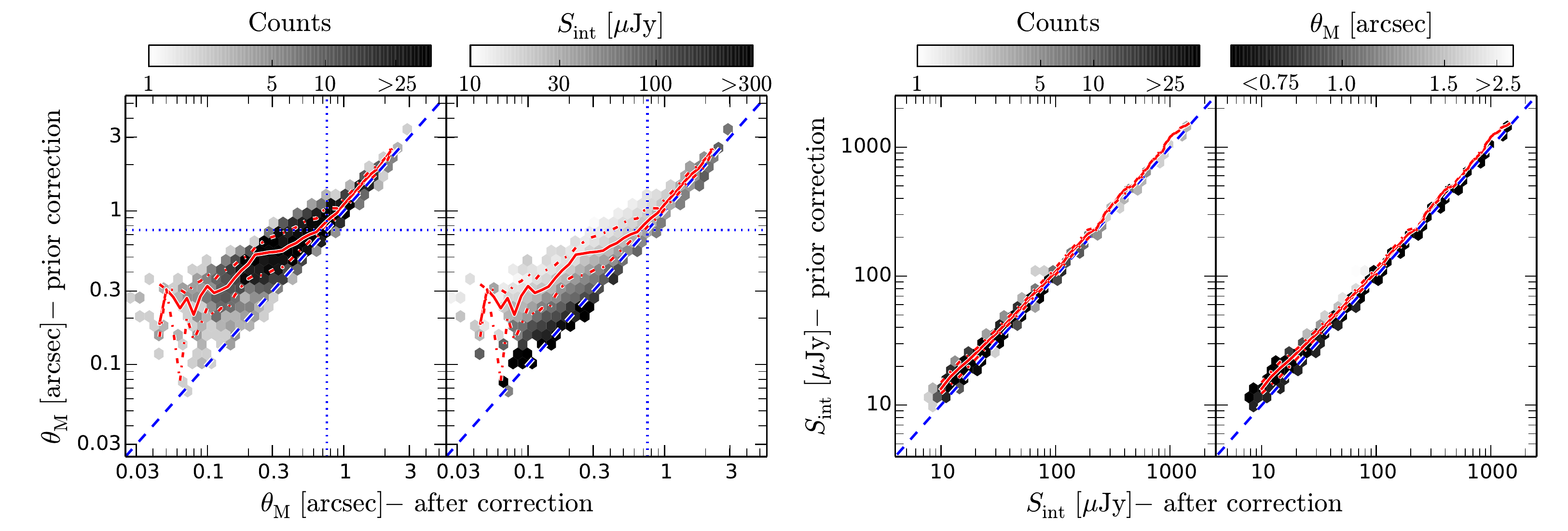}
	\caption{ Comparison between  observed and corrected source parameters of SFGs in the sample. {\it Left panels:} FWHM of 2519 resolved sources before and after correction color-coded by number counts and median flux density. The red solid (dashed) lines show the 50th (16th, 84th) percentile of the FWHM values prior correction, using a 0.05 dex bin width along the x-axis. The dotted blue lines illustrate the FWHM of the synthesized beam (0.75\,arcsec), while the 1:1 relation is shown by the blue dashed line.  {\it Right panels:} flux density of 3184 SFGs  (both resolved and unresolved) before and after correction color-coded by number counts and median FWHM. The red solid (dashed) lines show the 50th (16th, 84th) percentile of the flux density values prior correction, using a 0.1 dex bin width along the x-axis. The 1:1 relation is shown by the blue dashed line.   }
	\label{fig:comparison}%
\end{figure*}

\item  After having applied our corrections to all mock resolved sources,  we  retrieve the  distribution of $r_{\theta}$ and $r_{S}$. By fitting a single Gaussian component, we find  $\mu=0.0$ and $\sigma=1.0$ for both distributions (Fig. \ref{fig:sigmadistributions}).  This assures that the corrected  flux densities and { FWHM}, as well as their associated uncertainties, are a good description of the input mock sources.\\

Note that for a small fraction of mock sources, our corrected FWHM is still being underestimated;  giving rise to a wing in the $r_{\theta}$ distribution (Fig. \ref{fig:sigmadistributions}). 
We verified that  these outliers are mainly located at the  extended and bright-end of the $\theta_{\rm M}^{\rm out} - S_{\rm int}^{\rm out}$ plane ($\theta_{\rm M}^{\rm out} > 0.75\rm \, arcsec$ and   $S_{\rm int}^{\rm out} > 0.1\rm mJy$), where less than 1\% of SFGs in our final sample reside (see Fig. \ref{fig:completeness}). \\

\item  To correct the measured flux density of unresolved sources, we compare the input and output flux density of mock sources retrieved as unresolved by {\tt PyBDSF}  (see Fig. \ref{fig:fluxboosting_unresolved}). We then derive ``flux boosting''  factors as a function of S/N; while at $\rm S/N = 5$ the flux density is overestimated by 17\%, at $\rm S/N> 7$ the effect of ``flux boosting'' is negligible. \\

\item We verified that the corrections, as well as the completeness, do not strongly depend on the input angular size and flux density distribution used in the MC simulations. A uniform distribution (equal number of sources per bin in the $\theta_{\rm M}^{\rm in} - S_{\rm int}^{\rm in}$ plane) yields correction factors that are consistent with those obtained from a realistic input distribution.

\end{enumerate}

Having validated our method, we then derive the corrected flux density  and size of SFGs in our sample. In Fig. \ref{fig:comparison}, we compare the flux and FWHM before and after revision in order to  illustrate the  effect of our corrections. Both flux and size measurements appear to be overestimated for faint radio sources. This result is expected, as positive noise fluctuations enhance the flux density on a pixel-by-pixel basis and, consequently, the amplitude and variance of a 2D Gaussian model are magnified. This phenomenon translates into a ``flux boosting'' factor of \hbox{$\sim20\%$} at the faint-end (see right panel of Fig. \ref{fig:comparison}) -- comparable with the uncertainty for the flux density of a 5$\sigma$ radio source detection.  
On the other hand, ``size boosting'' seems to be ubiquitous for faint and compact sources, that have a deconvolved FWHM smaller than the size of the synthesized beam  (see left panel of Fig. \ref{fig:comparison}). This can be attributed to the large uncertainties associated with the deconvolution process of slightly resolved and faint radio sources.   \\

As a consistency test,  we compare our corrected flux density measurements with those reported by \citet{smolcic17b}, which were  derived following a non-parametric approach with {\tt blobcat}. By considering both resolved and unresolved sources  (see Fig. \ref{fig:blobcatvspybdsf}), we found that both quantities are, on average,  consistent.

\subsection{From flux and size measurements to SFR and effective size estimates}  \label{subsec: sfrandsize}

We estimate the total SFR by adding the estimates from the infrared ($\rm SFR_{IR}$)  and uncorrected UV emission ($\rm SFR_{UV}$),  allowing us to account for the dust obscured and unobscured star formation activity.
We use the \cite{kennicutt98} calibration and the infrared-radio correlation \citep[e.g. ][]{magnelli15, delhaize17}  to derive $\rm SFR_{IR}$ as follows:
\begin{equation}
{\rm SFR_{IR} \, [M_\odot\,yr^{-1}]}= f_{\rm IMF}10^{-24}10^{q_{\rm IR}}L_{\rm 1.4GHz} \, [\rm W\,Hz^{-1}],
\end{equation} 
\noindent
where $f_{\rm IMF}=1.72$ for a Salpeter initial mass function (IMF) and $q_{\rm IR}$ is parametrized as a function of redshift (for SFGs only) as $q_{\rm IR}=(2.83\pm0.02)\times(1+z)^{-0.15\pm0.01}$  \citep{delhaize17}. $L_{\rm 1.4\,GHz}$, on the other hand, can be derived from the observer-frame 3 GHz fluxes $(S_{\nu _{3\,\rm GHz}}\, [\rm W\,Hz^{-1}\,m^{-2}])$  through:
\begin{equation}
L_{\rm1.4\,GHz}=\frac{4\pi D_L(z)^2}{(1+z)^{1-\alpha}}\left(\frac{1.4\,}{3}\right)^{-\alpha}S_{3\,\rm GHz},
\end{equation}
\noindent
where $D_{\rm L}$ is the luminosity distance in meters and $\alpha$ is the spectral index of the synchrotron power law  ($S_\nu\propto \nu^{-\alpha}$) of 0.8 \citep{condon92}. \\

We use in addition  the near-UV (NUV) emission of galaxies, from the COSMOS2015 catalog \citep{laigle16}, to estimate $\rm SFR_{UV}$ as follows \citep{kennicutt11}:
\begin{equation}
{\rm SFR_{UV}}= 10^{-43.17} L_{\rm NUV} \, [\rm erg\, s^{-1}]. 
\end{equation}

\begin{figure*}
	\centering
	\includegraphics[width=18.2cm]{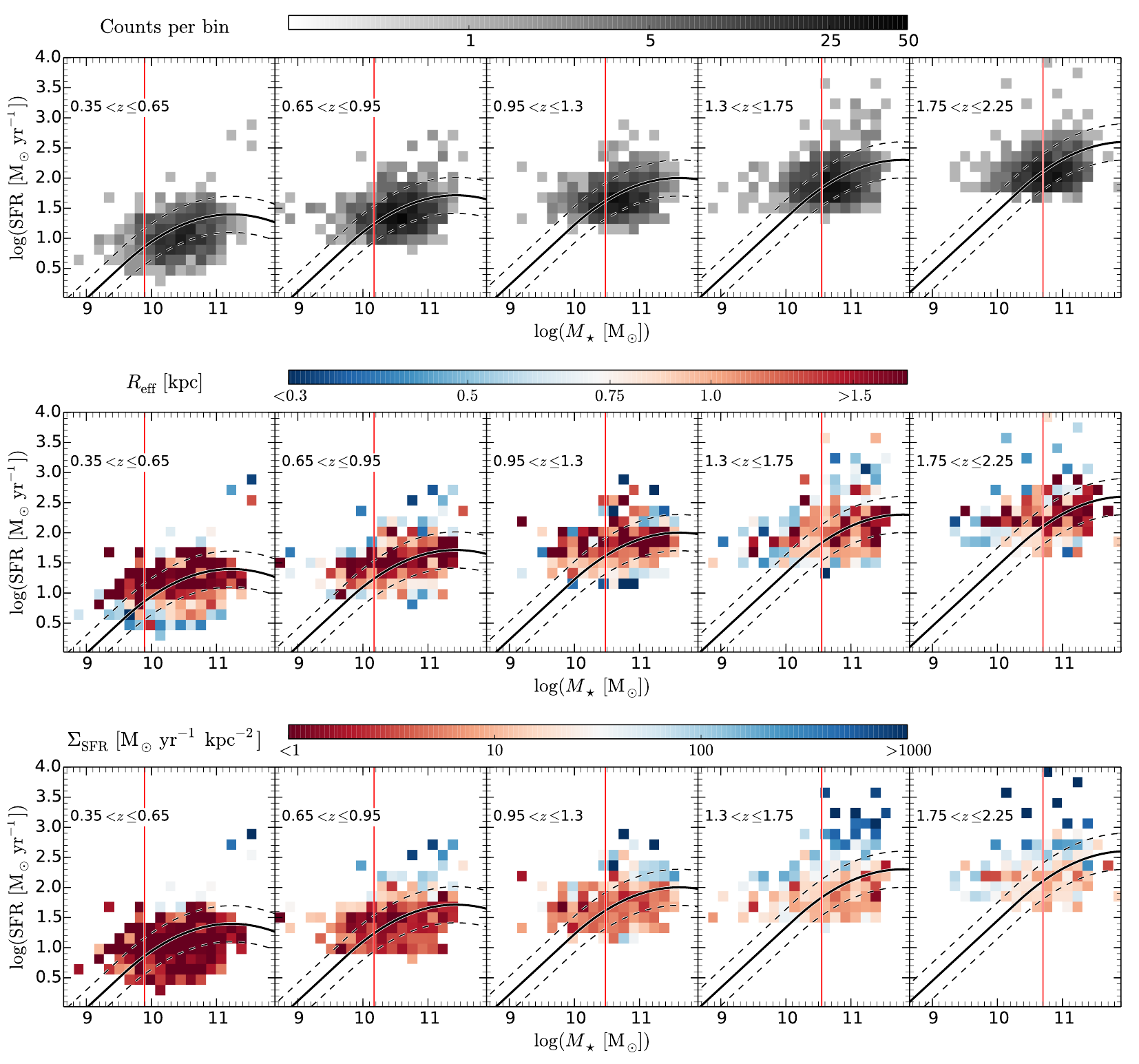}
	\caption{ {\it Upper panels:} 3184 SFGs in the ${\rm SFR}-M_{\star}-z$ plane.  Color scale shows the number of sources per bin. Solid/dashed black lines show the position of the MS of SFGs and its associated dispersion given by \citet{schreiber15}. Vertical red solid lines show the mass-limit above which we consistently probe SFGs on and above the MS. {\it Middle and lower  panels:} median size  and star formation surface density, respectively, of 3184 SFGs in the sample. We use 1000 MC realizations to include the values for unresolved sources, which are drawn from the distributions described in Sects. \ref{subsec:size-dms}   and  \ref{subsec:surfacedensity}.  }
		 
	\label{fig:sfr-mass-z_plane}%
\end{figure*}

Lastly, to compare our radio continuum  size estimates with those derived from the  optical/UV,  we  convert our $\theta_M$ measurements into effective radius ($R_{\rm eff}$),  i.e., the radius enclosing half of the total flux density.   To this end, we assume that most of our galaxies  are star-forming disks with an exponentially declining surface brightness distribution. This is consistent with the average S\'ersic index of $n\sim1$ for MS galaxies \citep[e.g., ][]{nelson16b} and luminous sub-mm selected galaxies \citep[SMGs; ][]{hodge16},  preferentially located above the MS. Under this assumption,   \cite{murphy17} have analytically proven that  for slightly resolved radio sources (with $R_{\rm eff}\lesssim\theta_{\rm beam}$) $\theta_M$  and  $R_{\rm eff}$  can be related by  
\begin{equation}
\theta_M\approx 2.430 R_{\rm eff}.
\end{equation}

\subsection{Final sample}\label{subsec:finalsample}

We distribute the 3184  SFGs in our sample  in five redshift bins following those presented by \cite{laigle16}, i.e.  (0.35, 0.65], (0.65, 0.9], (0.9, 1.35], (1.35, 1.7] and  (1.75, 2.25]. This allows us to directly use the stellar mass completeness  limits (per redshift bin) of the COSMOS2015 catalog, and hence assemble a mass-complete sample of radio-selected SFGs in the COSMOS field. The number of SFGs per redshift bin is nearly homogeneous (with a median of $\sim650$ sources). Given the small comoving volume probed by COSMOS at low-redshift and the  selection function that restricts our parameter space to compact starburst galaxies,  we are not able to explore the size evolution of SFGs  in the redshift regime below $z=0.35$.

\begin{figure*}
	\centering
	\includegraphics[width=18.0cm]{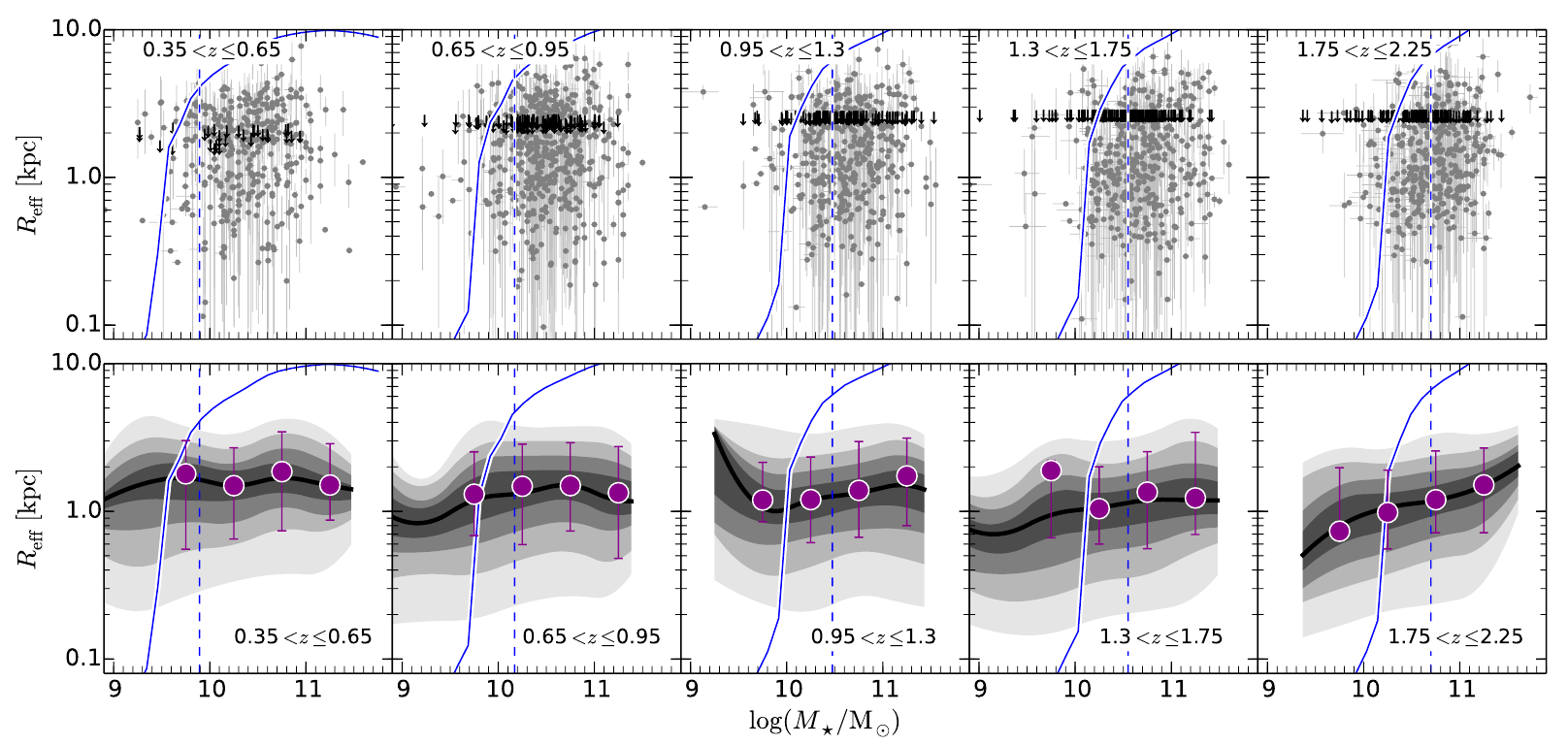} 
	\caption{ {\it Upper panels:}  3184 SFGs in the $R_{\rm eff}-M_{\star}-z$ plane. 
		The blue solid line shows the  maximum size, corresponding to the 10\% completeness level,  that can be observed for a galaxy with $\Delta \log(\rm SSFR)=0$  evaluated at the central redshift value per bin. 
		Blue dashed lines show the mass-limit above which we consistently probe SFGs on and above the MS. Black arrows show the upper limits for the size of unresolved sources. {\it Lower panels}: density distribution per stellar mass bin (0.5 dex width) of SFGs in the  $R_{\rm eff}-M_{\star}-z$ plane. Contour levels are at the 10, 20, 30, 40, 50, 60, 70, 80 and 90th  percentile. The median size derived via the KM estimator is shown by the dark magenta circles, the error bars correspond to the 16th and 86th percentile of the distribution.  }
	\label{fig:mass-size}
\end{figure*}

In Fig. \ref{fig:sfr-mass-z_plane}, we present the sample of 3184 SFGs in the SFR$-M_\star-z$ plane. The bulk of the radio-selected SFGs is consistent with the position and dispersion of the MS of SFGs, as given by \citet{schreiber15}. At the low mass-end, however, our radio detection limit biases our sample towards the starburst population. Since we aim to statistically analyze the size distribution of SFGs on and above the MS, we need to focus on the high-mass end. For this purpose, we define a mass-limit ($M_\star^{\rm lim}$) for each redshift bin, above which we are able to consistently probe both SFGs on ($-0.3\leq\Delta\log(\rm SSFR)_{\rm MS} \leq 0.3$)  and above the MS ($\Delta \log(\rm SSFR)_{\rm MS} > 0.3$). By considering systems with  $M_\star>M_\star^{\rm lim}$  we are also able to assemble a mass-complete sample of radio-selected SFGs, given that -- in all redshift bins --   $M_\star^{\rm lim}$  is  higher than  the stellar mass completeness limit of the COSMOS2015 catalog.

\section{Results}\label{sec:results}

In this section, we explore the dependence of the radio continuum size ($R_{\rm eff}$) on the stellar mass, distance to the MS and redshift.  We carefully address these relations while keeping in mind the completeness/size biases mentioned in Sec. \ref{subsec:mcsimulations} and  that our analysis is restricted to $M_\star>M_\star^{\rm lim}$, i.e. the part of the parameter space where the sample of SFGs on and above the MS is complete.  We also verified that the trends presented below remain even if we  use non-corrected measurements (see Appendix \ref{appendix:b}).  

\subsection{Radio continuum size {\it vs} stellar mass}\label{subsec:size-mass}

The stellar mass-size relation in galaxies \citep[e.g., ][]{furlong17, allen17} is thought to be linked to the physical processes that regulate galaxy assembly, such as galaxy minor/major mergers and gas accretion \citep[e.g., ][]{khochfar06, khochfar09, dekel09, oser10, gomez-guijarro18}.  Thus, it is a fundamental ingredient to understand galaxy  evolution. 

Here, we attempt to characterize the stellar mass-radio size relation up to $z=2.25$.  We thus explore the scatter of SFGs in the $R_{\rm eff}- M_{\star}$ plane by deriving their density distribution per stellar mass bin ($0.5\,\rm dex$ width; Fig. \ref{fig:mass-size}). We  use 10,000 Monte Carlo trial model runs to take into account the dispersion introduced by the uncertainties and upper limits of $R_{\rm eff}$ for  resolved and unresolved sources, respectively. Based on our MC simulations (Fig. \ref{fig:upperlimitunresolved}),  the size of unresolved sources can be drawn from  a uniform distribution - in log space - within the range $ [0.1, R_{\rm eff}^{\rm lim}]\, \rm kpc$, where $R_{\rm eff}^{\rm lim}$ is the upper limit for the source size. We  also derived the median size of SFGs    through the Kaplan–Meier (KM) estimator \citep{kaplan58}, which allows us to take into account the upper limits for the size of unresolved sources.   We find that both methods, MC realizations and KM estimator, yield consistent results (Fig. \ref{fig:mass-size}). In all redshift bins, the size distribution of SFGs remains constant over the range of stellar mass probed here, where the median size differs in less than 25\% (see Table \ref{table:1}). Qualitatively, this result is  consistent with the shallow slope ($\alpha_{\rm opt/UV}$) of  the stellar mass and optical/UV size  relation of SFGs  \citep[$\alpha_{\rm opt/UV}\sim0.2$; e.g., ][]{vanderwel14, mowla18}.  Lastly, we  checked that this relation remains   if we use two separate samples of SFGs: one composed by galaxies on the MS ($-0.3\leq \Delta\log(\rm SSFR)_{\rm MS}\leq 0.3$) and another  above it ($\Delta\log(\rm SSFR)_{\rm MS}>0.3$).

We still have to consider that the latter result might be affected by our selection function.  As  mentioned in Sect. \ref{subsec:mcsimulations},  galaxies are preferentially detected if they are compact, especially at the faint-end.  This could yield a misleading stellar mass - radio size relation, as low-mass SFGs are fainter than their massive counterparts (due to the MS slope, Fig. \ref{fig:sfr-mass-z_plane}). To quantify this possible bias, we use our MC simulation's output to estimate the maximum recoverable size as a function of stellar mass as follows. At a given redshift bin and for each mass, we infer the SFR of a galaxy with $\Delta\log(\rm SSFR)_{\rm MS}=0$. Then we convert this  SFR into flux density using the central redshift of the bin. Finally, this flux is associated to a maximum recoverable size using  our 10\% completeness limit in the $\theta_{\rm M}^{\rm in}-S_{\rm int}^{\rm in}$ plane (Sect.  \ref{subsubsec:selectionfunction}). As observed in Fig. \ref{fig:mass-size}, this selection function  hinders the detection of extended SFGs with  stellar mass below and near $M_{\star}^{\rm lim}$; yet it does not affect the parameter space above $\log(M_{\star}/\rm M_\odot)=10.5$. 
 Hence, the negligible dependence  of the stellar mass on the radio size of SFGs with $\log(M_{\star}/\rm M_\odot)>10.5$  remains unaffected by our selection.

\begin{figure*}
	\centering
    \includegraphics[width=18.0cm]{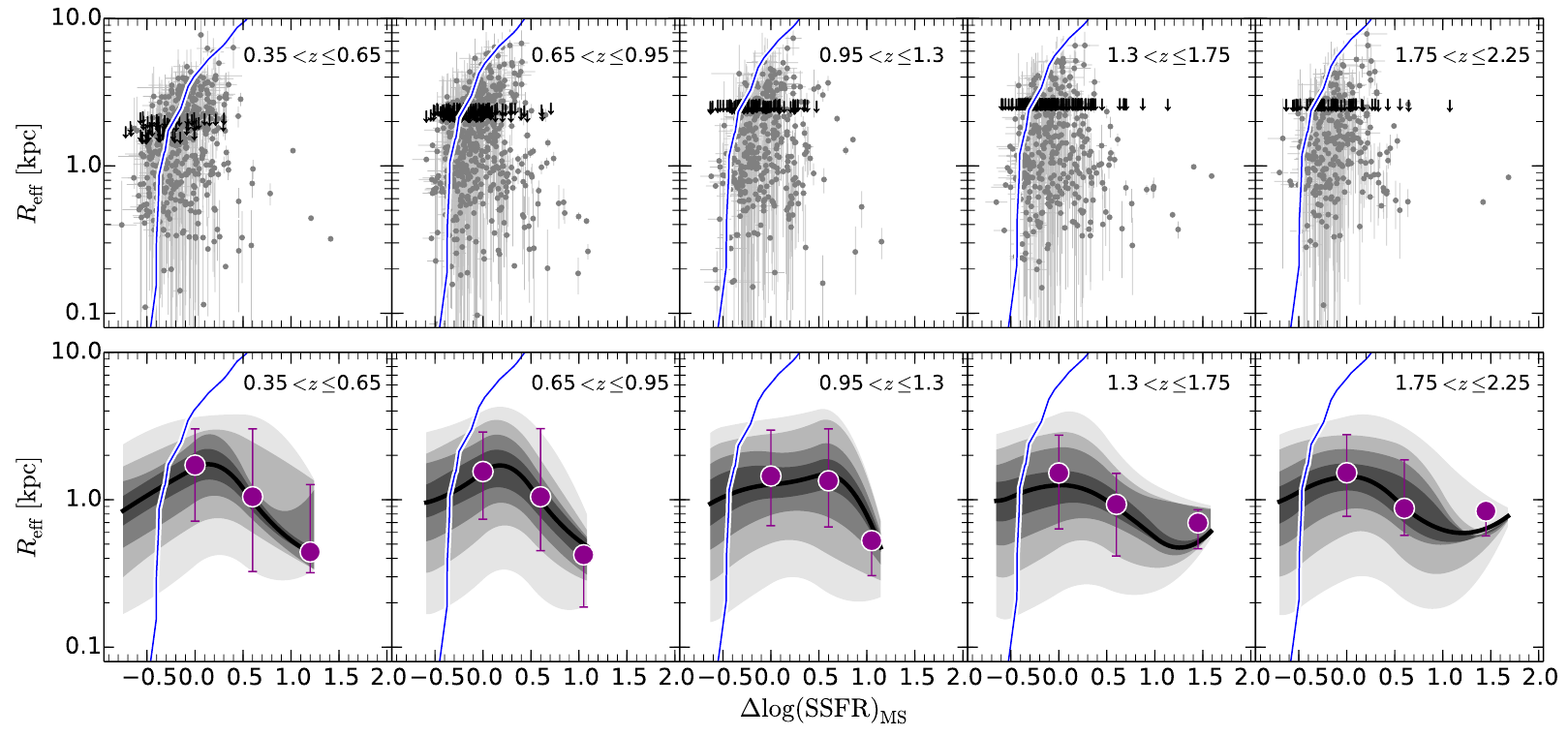}   
	\caption{ {\it Upper panels:} SFGs with $M_\star>M_\star^{\rm lim}$  in the size$-\Delta\log({\rm SSFR})-z$ plane. 	The blue solid line shows the  maximum size, corresponding to the 10\% completeness level,   that can be observed for a galaxy with $M_\star=M_\star^{\rm lim}$ evaluated at the central redshift value per bin.  Black arrows show the upper limits for the size of unresolved sources.   {\it Lower panels:} density distribution per $\Delta\log(\rm SSFR)$ bin (0.5 dex width) of SFGs in the  size$-\Delta\log({\rm SSFR})-z$ plane. Contour levels are at the 10, 20, 30, 40, 50, 60, 70, 80 and 90th  percentile. The median size derived via the KM estimator is shown by the dark magenta circles, the error bars correspond to the 16th and 86th percentile of the distribution. } 
	\label{fig:size-dms}%
\end{figure*}

\subsection{Radio continuum size of SFGs on and above the main-sequence}\label{subsec:size-dms}

Since both the size and $\Delta \log( \rm SSFR)_{\rm MS}$ of SFGs can be discussed within the context of gas accretion and merger-driven star formation \citep[e.g., ][Lang et al. in prep.]{elbaz11, elbaz18}, it is essential to  characterize their interplay in detail. We therefore take  advantage of our mass-complete sample of radio-selected SFGs to systematically explore their size distribution as a function of $\Delta \log( \rm SSFR)_{\rm MS}$  and cosmic time  (Fig. \ref{fig:size-dms}). We remind the reader that we consider SFGs with  $M_\star>M_\star^{\rm lim}$, that is the region of the parameter space where we can consistently probe galaxies on and above the MS.

Similarly to the previous section, we derive the median size of SFGs per $\Delta \log (\rm SSFR)_{\rm MS}$  bin following a MC approach and using the KM estimator.  As observed in Fig. \ref{fig:size-dms}, both methods agree well and reveal a trend where SFGs with higher $\Delta\log(\rm SSFR)_{\rm MS}$ values are compact; in particular at  lower redshifts where this tendency is more pronounced.  The median size of $z\sim0.5\,\, (z\sim2)$ MS galaxies is 4 (2) times larger than those with $\Delta\log(\rm SSFR)_{\rm MS}>0.9$ (see Table \ref{table:2}). Note that  although SFGs on the MS are preferentially extended \hbox{(median $R_{\rm eff}\sim1.5\pm0.2\,\rm kpc)$}, some  can be as  compact as galaxies with elevated $\Delta\log(\rm SSFR)_{\rm MS}$.

To verify that our selection function does not bias these trends, we   estimate the maximum recoverable size as a function of $\Delta \log( \rm SSFR)_{\rm MS}$. At a given redshift bin and for each $\Delta \log( \rm SSFR)_{\rm MS}$, we infer the SFR of a galaxy with $M_\star=M_\star^{\rm lim}$ evaluated at the central redshift of the bin. We then convert this SFR into flux density and associate it to a maximum recoverable size  using our completeness in the $\theta_{\rm M}^{\rm in}-S_{\rm int}^{\rm in}$ plane (Sect. \ref{subsubsec:selectionfunction}).  As inferred from Fig. \ref{fig:size-dms}, while the  selection function  does not affect the size distribution of SFGs above the MS, it does   hamper the detection of  extended SFGs on and below the MS.   We emphasize that retrieving these missing systems would only strengthen the anti-correlation between the median size of SFGs and  $\Delta \log( \rm SSFR)_{\rm MS}$.

The size dichotomy of SFGs on and above the MS  is consistent with the results of   \cite{elbaz11} and \cite{rujopakarn16}.  They did not report, however, that  compact galaxies can be ``hidden'' within the MS (Fig. \ref{fig:size-dms}), which is in agreement with recent high-resolution ALMA observations of $z\sim2$ SFGs \citep[e.g., ][Lang et al. in prep.]{elbaz18}.  The latter is related to the stacking approach followed by \cite{elbaz11},  that can only provide  median quantities  and  the small sample of \cite{rujopakarn16} that  might be affected by incompleteness.  
 
In general, these findings  support the emerging consensus where  the global star-forming region of SFGs on the MS  is preferentially but not exclusively extended, while  SFGs above the MS are always more compact systems.

\subsection{Size of SFGs in different wavelengths and its evolution with redshift}\label{subsec:sizecomparison}

\begin{figure}[h!]
	\centering
	\includegraphics[width=8.9cm]{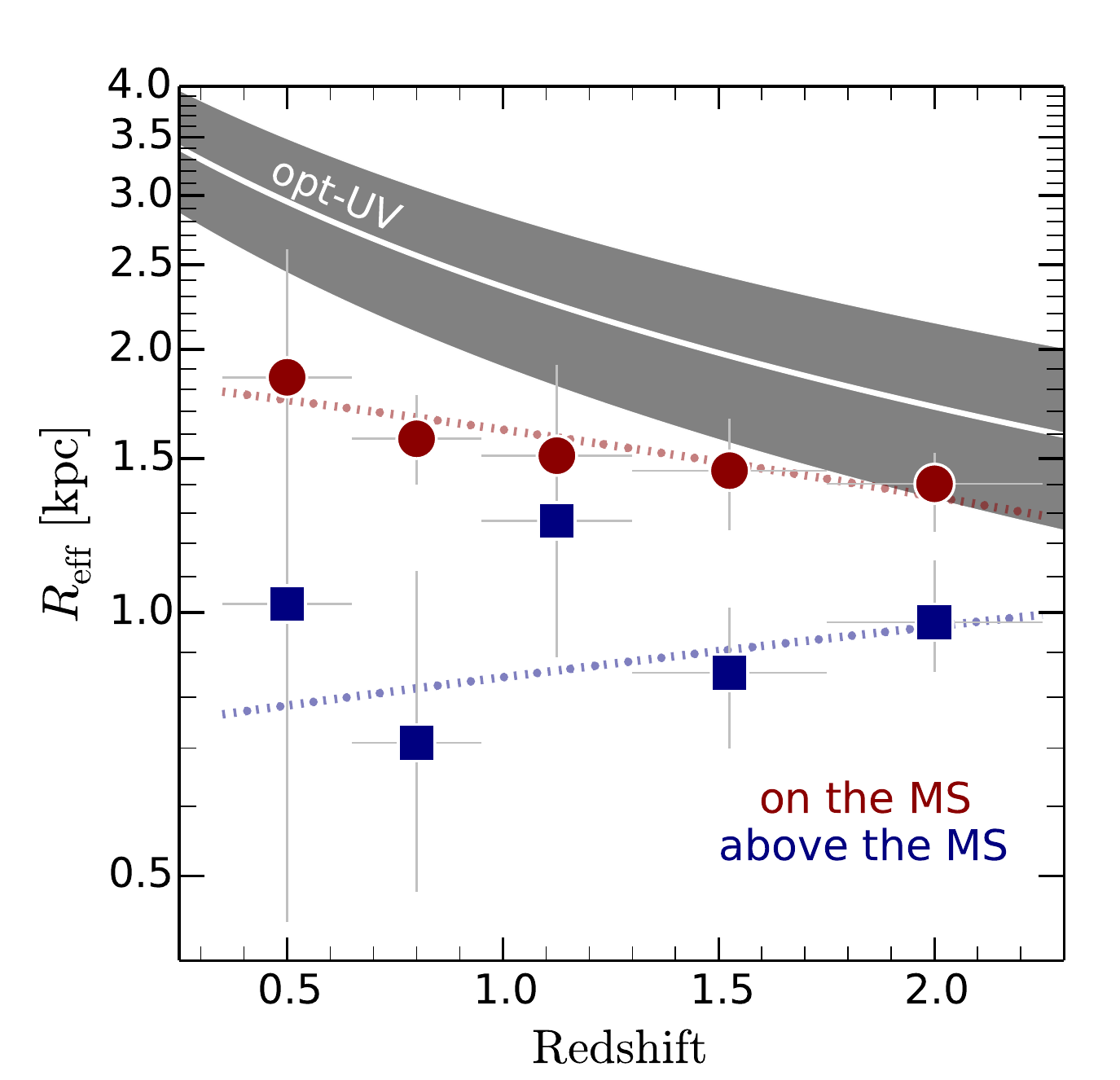}
	\caption{Radio continuum size (in terms of half-light radius, $R_{\rm eff}$) of galaxies  on and above the  MS  as a function of redshift. Only SFGs with  $\log(M_\star/\rm M_\odot)>10.5$ are included. Filled data points (squares/circles) show the median size for SFGs (above/on the MS) in the different redshift bins probed in this work. Vertical bars show the 95\% interval confidence of the median, while horizontal bars represent the redshift bin width. 	Grey shaded region illustrates the growth curve derived in the UV-optical for UV-luminous SFGs given by  $R_{\rm eff}/{\rm kpc}=(4.78\pm0.68)(1+z)^{(-0.84\pm0.11)}$ \citep{shibuya15}. Red and blue dotted lines show the linear fit to parametrize the redshift evolution of the median radio continuum size as: $R_{\rm eff}/{\rm kpc}=(2.1\pm0.2)(1+z)^{(-0.26\pm0.08)}$ and    $R_{\rm eff}/{\rm kpc}=(0.6\pm0.4)(1+z)^{(0.12\pm0.14)}$, for SFGs on and above the MS, respectively.  }
	\label{fig:size-redshift}%
\end{figure}

We now explore the radio continuum size evolution of  SFGs over the redshift range $0.35<z<2.25$ to better constrain the processes regulating the growth of galaxies. In addition,  through the comparison of the size-redshift relation as traced by stellar light, dust and supernovae remnant, we investigate where and how new stars are formed in galaxies. 
To this end, we select galaxies from our final sample (Sect. \ref{subsec:finalsample}) with $\log(M_\star/\rm M_\odot)>10.5$, which is the only mass bin consistently probed across the redshift range explored here. For all the redshift bins, we then derive the median $R_{\rm eff}$ (via the KM estimator) of SFGs on and above the MS, i.e. $-0.3\leq \Delta\log(\rm SSFR)_{\rm MS}\leq 0.3$) and  ($\Delta\log(\rm SSFR)_{\rm MS}>0.3$), respectively.  As illustrated in Fig. \ref{fig:size-redshift},   the radio continuum size of both SFG populations  remains nearly constant across cosmic time. By using a parametrization of the form $R_{\rm eff}\propto(1+z)^{\rm \alpha}$, we find a slope of only $-0.26\pm0.08$ ($0.12\pm0.14$)  for  SFGs on (above) the MS.  As expected from the results in Sect. \ref{subsec:size-dms}, the median size of SFGs on the MS ($R_{\rm eff}=1.5\pm0.2\rm$\,kpc) is significantly larger than for those above it ($R_{\rm eff}=1.0\pm0.2$\,kpc). \\

The  size evolution presented here is still influenced by 
two factors that can not be characterized with the available data: 
\begin{itemize}
	\item  Firstly,  our radio continuum size estimates trace different rest-frame frequencies, from 4\,GHz at $z=0.35$ to 9.7\,GHz at  $z=2.25$.   Since higher energy electrons lose energy more rapidly, their propagation throughout the galaxy is more limited than their low-energy counterparts \citep[e.g., ][]{kobayashi04}.  3 GHz emitting  electrons, in particular, are expected to diffuse $\sim25\%$ further into the ISM than those at 10 GHz \citep[][]{murphy17}.  Thus, the observed radio continuum synchrotron emission would tend to be more concentrated as the redshift increases. This phenomenon does not significantly affect the trends presented above,  as a 25\% larger radio size at $z=2.25$ would  lead to $\alpha\sim -0.10$ ($\sim0.25$) for  SFGs on (above) the MS. 

	\item Secondly, given the limited number of resolution elements across the SFGs in the sample,   we can not assess their radio continuum surface brightness distribution and determine $R_{\rm eff}$.  Therefore, 	 we converted the  deconvolved FWHM to $R_{\rm eff}$ \citep[following ][]{murphy17} assuming that most of our SFGs follow an exponentially declining surface brightness distribution (with S\'ersic index $n=1$). Naturally, such a conversion might deviate from the true $R_{\rm eff}$ on a galaxy-by-galaxy basis, specially for SFGs above the MS that tend to have cuspier light profiles \citep[e.g., ][]{wuyts11}. For slightly resolved sources, like the ones presented here, we do not expect major changes in $R_{\rm eff}$ if the  S\'ersic index is larger than 1. For example, even for  0.2\,arcsec resolution dust-continuum observations, \citet{elbaz18} report that both $R_{1/2}\equiv 0.5\times \rm FWHM$ and $R_{\rm eff}$ are an equally good proxy for the half-light radius, either leaving the index free or fixed to 1. 
	 
\end{itemize}

\subsubsection{Comparison with other radio continuum size estimates}\label{subsubsec:sizecomparisonradio}

\cite{bondi18} have independently derived $R_{\rm eff}$ of radio sources detected in the VLA COSMOS 3GHz map \citep{smolcic17}, including AGN and SFGs  up to $z\sim3$. They assembled a sample of SFGs that is complete in $L_{3\rm GHz}$ over $0.8<z<3$ with  median $\log(M_\star/\rm M_\odot)=10.6$, no distinction was made between on and off MS galaxies. For this sample, the size of SFGs marginally increases with cosmic time: from median \hbox{$\sim 1.4$\,kpc}  at  $z=2.1$ to \hbox{$R_{\rm eff}\sim 1.6$\,kpc} at $z=0.8$. This is in agreement with the shallow size evolution of MS galaxies in our sample (which correspond to $\sim 90\%$ of all SFGs) with median $R_{\rm eff}=1.5\pm0.1$\,kpc,  within the same redshift range and comparable stellar mass. We note that despite the  independent methodologies used to measure radio sizes  in the $\mu\rm Jy$ regime, and  different selection criteria, our median sizes are  consistent.  \cite{bondi18} used, in particular,  the original and convolved images (up to a resolution of 2.2 arcsec) of the VLA COSMOS 3GHz mosaic and took the flux density provided by {\tt blobcat} as a prior in their 2D Gaussian fitting procedure. This flux prior limits the effect of ``size boosting'',  leading to comparable size measurement between our two studies.  On the other hand, by using the VLA COSMOS 3GHz map, \citet{miettinen17} reported a median  $R_{\rm eff}\sim1.9\rm \, kpc$ for  152 SMGs over the redshift range of $1\lesssim  z \lesssim 6$.   The discrepancy of $\sim35\%$ between the latter $R_{\rm eff}$ and that reported here at $z=2.25$, and in \citet{bondi18}, is likely driven by the different selection criteria.

The angular size of the $\mu\rm Jy$ radio sources has also been recently explored in different extragalactic deep fields. At the same  frequency, 3\,GHz, it was found that \hbox{$z\sim1$}   SFGs in the Lockman-Hole have a median effective radius of  \hbox{$\sim1.0 \rm\,kpc$} \citep{cotton18}, which agrees  with the median size of SFGs above the MS derived here (see Fig. \ref{fig:size-redshift}). \cite{murphy17} have  reported that $z\sim1.2$ GOODS-N SFGs  have a median $R_{\rm eff}$ of only  $\sim0.5\,\rm kpc$ at 10\,GHz. This small size could be associated to selection effects, as the high resolution  10\,GHz observations  (0.22\,arcsec) are sensitive to smaller angular scales.  Additionally, as stated by \citet{murphy17}, the discrepancy between 3\,GHz and 10\,GHz radio sizes is driven by the frequency-dependent  cosmic ray diffusion.  This physical phenomenon could partially explain the large median $R_{\rm eff}$ of $2.3\pm0.6 \rm\,kpc$ at 1.4\,GHz of $z\sim1.5$ SFGs \citep[in the Hubble Deep Field; ][]{lindroos18}, which is $\sim1.5$ times larger than the median size at 3\,GHz of galaxies in our sample. The larger energy loss rate at higher frequencies can not explain, on the contrary, the  large  median $R_{\rm eff}$ of $\sim 2.7 \rm\,kpc$ ($\rm FWHM\sim0.8$ arcsec) at 5.5\,GHz  reported  for SFGs at similar redshift  \citep[in GOODS-N; ][]{guidetti17}. 
 Lastly, we note that 
 (apart from the  frequency and resolution of the observations) a more general issue about size determination is  related to the surface brightness limit of each survey. As inferred from Fig. \ref{fig:completeness}, a lower r.m.s sensitivity  hampers the maximum detectable angular size, biasing the sample towards more compact radio sources.

\subsubsection{Comparison with FIR, optical and $\rm H\alpha$ sizes }\label{subsubsec:sizecomparison_others}

\begin{figure*}
	\centering
	\includegraphics[width=18cm]{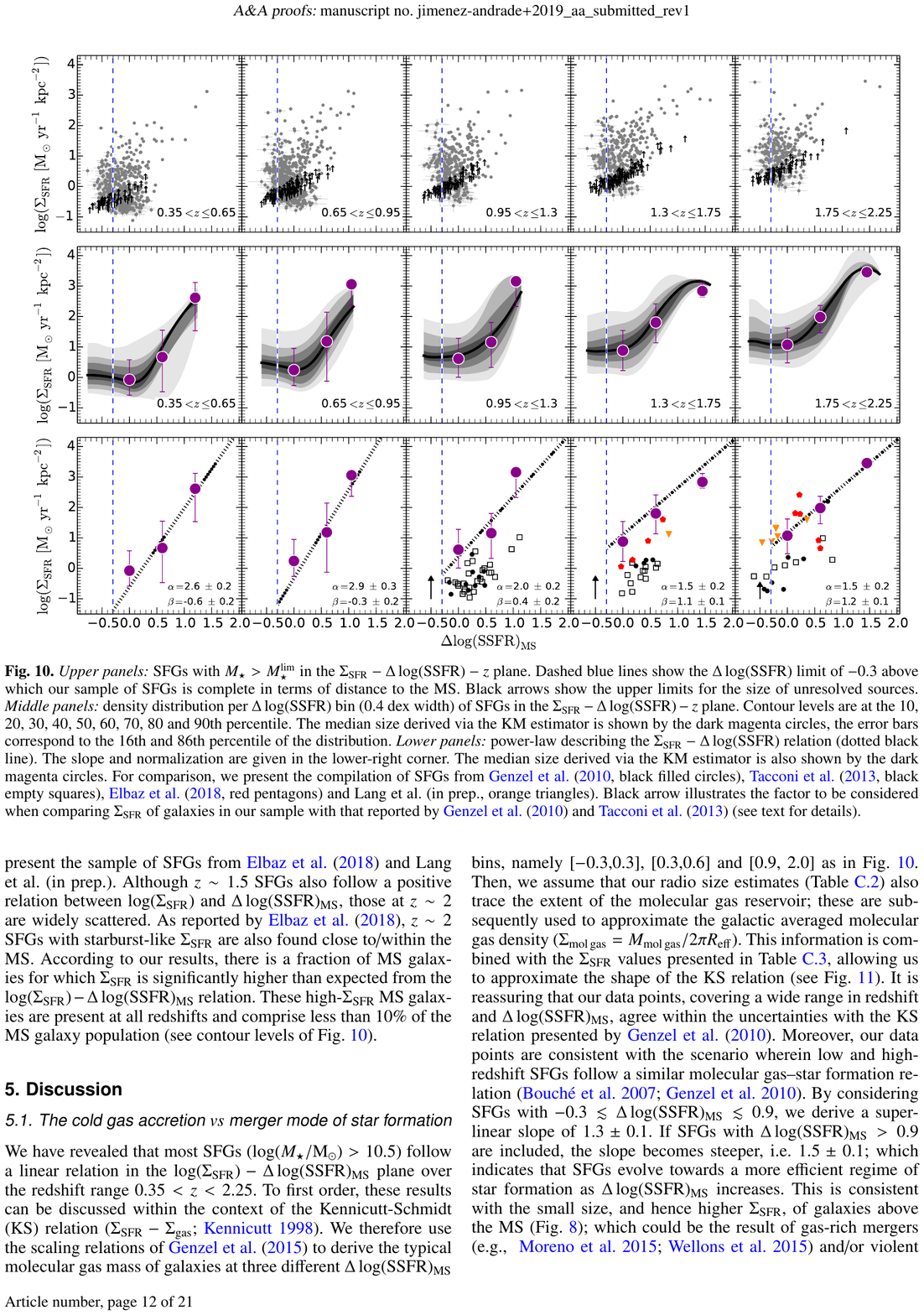}
	\caption{{\it Upper panels:}  SFGs with $M_\star>M_\star^{\rm lim}$  in the $\Sigma_{\rm SFR}-\Delta\log({\rm SSFR})-z$ plane. 	Dashed blue lines show the $\Delta\log(\rm SSFR)$ limit of $-0.3$ above which our sample of SFGs is complete in terms of distance to the MS. Black arrows show the upper limits for the size of unresolved sources.   {\it Middle panels:} density distribution per $\Delta\log(\rm SSFR)$ bin (0.4 dex width) of SFGs in the  $\Sigma_{\rm SFR}-\Delta\log({\rm SSFR})-z$ plane. Contour levels are at the 10, 20, 30, 40, 50, 60, 70, 80 and 90th  percentile. The median size derived via the KM estimator is shown by the dark magenta circles, the error bars correspond to the 16th and 86th percentile of the distribution.   {\it Lower panels:} power-law describing the $\Sigma_{\rm SFR}-\Delta\log(\rm SSFR)$ relation (dotted black line). The slope and normalization are given in the lower-right corner. The median size derived via the KM estimator is also shown by the dark magenta circles. For comparison, we present the compilation of SFGs from \citet[][black filled circles]{genzel10}, \citet[][black empty squares]{tacconi13},   \citet[][red pentagons]{elbaz18} and Lang et al. (in prep., orange triangles).  Black arrow illustrates the factor to be considered when comparing  $\Sigma_{\rm SFR}$ of galaxies in our sample with that reported by \cite{genzel10} and \cite{tacconi13} (see text for details).  }
	\label{fig:surfacedensity}%
\end{figure*}

It has been reported that the FIR size of $z\sim2$ SFGs is, in average, $\sim1.5$\,kpc \citep[e.g., ][Lang et al. in prep.]{ rujopakarn16, elbaz18}, which is consistent with the median radio size of SFGs reported here. 
Extinction-corrected $\rm H\alpha$ radial profiles tracing the global star-forming region  of $z\sim1.4$ SFGs yield a median effective radius  of $<1$\,kpc \citep[with  $9.8<\log(M_\star/{\rm M_\odot})<11$; ][]{nelson16}, comparable with the radio continuum sizes of SFGs above the MS. In contrast, the median  effective radius derived from uncorrected $\rm H\alpha$ emission is  $4.2\pm0.1$\,kpc -- for  SFGs at $z\sim1$ and similar stellar mass  \citep[][]{nelson12}. Such disparity is  related to the high dust content in massive galaxies; if star formation is highly obscured at small radii,  $\rm H\alpha$  emission would appear less centrally concentrated and hence the inferred effective radius will be larger \citep[e.g., ][]{mollenhoff06,nelson16, nelson16b}.

In Fig. \ref{fig:size-redshift}, we also present a comparison of the size of SFGs as observed from radio continuum and   optical-to-UV throughout cosmic time. We use the size evolution derived by \cite{shibuya15}  via broadband {\it HST} ACS and WFC3 imaging. In this case, we adopt the median  fit obtained for the UV bright galaxies ($-24<M_{UV}<-21$)  corresponding to the stellar mass range of $10<\log(M_\star/\rm M_\odot)<11$, which is consistent with  the mass range of SFGs used in this work. Given  that \cite{shibuya15} masked star-forming clumps, their size estimates  can be used as a proxy for the stellar mass distribution  of galaxies. As revealed by Fig. \ref{fig:size-redshift}, the overall star-forming region of SFGs (on and above the MS) is more compact than their stellar component.  In particular, at $z=0.5$ (2) the optical-to-UV emission is  $\sim$2 (1.3) times more extended  than the radio continuum.  We note that, since those  {\it HST}/CANDELS-based estimates are not corrected for extinction,  it is likely that (similar to what has been shown for $\rm H\alpha$  sizes) the optical-to-UV effective radius is  overestimated. Given that dust extinction becomes substantial in high-redshift galaxies \citep[e.g., ][]{leslie18}, we would expect that their  optical-to-UV size is overestimated by a larger fraction than those galaxies at lower redshifts. Correcting for this effect could alleviate the discrepancy between the extent of the stellar and star-forming component of  galaxies at high redshift.

In summary,  radio continuum, FIR and extinction-corrected  $\rm H\alpha$ emission  suggest that star formation occurs in smaller regions relative to the total stellar component \citep[e.g., ][Karoumpis et al. in prep.]{simpson15, rujopakarn16, fujimoto17, elbaz18}.  Here, in particular, we find that while the radio continuum size remains nearly constant, the one inferred from optical-to-UV emission  increases with cosmic time.   In Sect. \ref{subsec:discussion_bulge}, we further discuss this finding within  the context of bulge growth.

\subsection{Cosmic evolution of $\Sigma_{\rm SFR}$ }\label{subsec:surfacedensity}

From numerical simulations, SFGs are expected to experience  a  compaction phase while the SFR increases and they move towards the upper end of the MS \citep[e.g., ][]{tacchella16}.  This scenario can now be tested through our radio continuum size estimates.  We thus derive the galactic-average SFR surface density, $\Sigma_{\rm SFR}={\rm SFR}/2\pi R_{\rm eff}^2$, and use it to evaluate how concentrated the star formation activity in galaxies is as they evolve across the MS (see Fig. \ref{fig:sfr-mass-z_plane}, \ref{fig:surfacedensity}). In calculating $\Sigma_{\rm SFR}$ we  assume that the total (UV/obscured+IR/unobscured) SFR is confined within the radio continuum-based $R_{\rm eff}$.  This simplification is valid as the UV/obscured SFR is considerable  small for massive, high-redshift SFGs \citep[e.g.,][]{buat12}; therefore, the star formation activity is  mainly traced by the radio continuum (unobscured) emission. 

As done in Sects. \ref{subsec:size-mass} and \ref{subsec:size-dms}, we follow a MC approach to derive the distribution of $\Sigma_{\rm SFR}$ per  $\Delta\log(\rm SSFR)_{\rm MS}$  bin (Fig. \ref{fig:surfacedensity}) -- again, only galaxies with $M_\star>M_\star^{\rm lim}$ are included in the analysis.  In this case, the $\Sigma_{\rm SFR}$ for unresolved sources  is drawn from a uniform logarithmic distribution  within the range [$\Sigma_{\rm SFR}(R_{\rm lim})$, $\Sigma_{\rm SFR}(\rm 0.1\,kpc)$].  We verify the reliability of this approach by using the KM estimator (Table \ref{table:3}). At all redshifts and for both methods, there is a positive relation between   $\Sigma_{\rm SFR}$ and $\Delta\log(\rm SSFR)_{\rm MS}$ that can be described with a power-law;
\begin{equation}
\log(\Sigma_{\rm SFR})=\alpha\times \Delta\log(\rm SSFR)_{\rm MS}+\beta,
\end{equation}
\noindent
where $\alpha\rm\, and\, \beta$ are the slope and normalization, respectively.
We  use a least-squares (Levenberg-Marquardt) algorithm to fit a linear function to $\Delta\log(\rm SSFR)_{\rm MS}$  and derive the best-fitting values for $\alpha\rm\, and\, \beta$. This procedure is done for each MC realization while restricting our parameter space to $\Delta\log(\rm SSFR)_{\rm MS}>-0.3$, where our sample is complete.  The final values for $\alpha\rm\, and\, \beta$, reported in Fig. \ref{fig:surfacedensity}, correspond to the median (and 16th and 84th percentile) after executing 1000 MC trial model runs. While the  normalization of the  $\log(\Sigma_{\rm SFR}) -  \Delta\log(\rm SSFR)_{\rm MS}$ relation  decreases with redshift,  the value of $\alpha$ reveals that this trend becomes shallower at higher redshift.  At $z\sim2$, the  difference between $\Sigma_{\rm SFR}$ of galaxies on and above the MS is smaller than in the local Universe.

Spatially-resolved studies of local  SFGs have also reported more centrally-peaked  radial  profiles of $\Sigma_{\rm SFR}$ as the distance to the MS increases \citep{ellison18}. It has been found that the FIR surface density evolves across the MS with a logarithmic slope of 2.6 \citep{lutz16}, which is  consistent with the value we have derived at $z\sim0.5$ ($\alpha=2.6$) using the radio continuum emission. Likewise,  from $\rm H\alpha$ resolved maps it has been shown that $z\sim1$ SFGs follow this relation with $\alpha\sim1.1$ \citep{magdis16}, which is $\sim$50\% lower than that reported in this work. This tendency can also be inferred from the $\Sigma_{\rm SFR}$, $M_\star$ and SFR of $1\lesssim z\lesssim 3$ SFGs reported by \cite{genzel10} and \cite{tacconi13}. As presented in Fig. \ref{fig:surfacedensity},  these SFGs also exhibit higher $\Sigma_{\rm SFR}$ as the distance to the MS increases, yet their $\Sigma_{\rm SFR}$  appear systematically lower than the values reported here. This could be a consequence of the optical/UV/H$\alpha$/CO-based size estimates used by the authors, which are larger than those obtained from radio continuum emission (Sect. \ref{subsubsec:sizecomparison_others}).  If we scale their $\Sigma_{\rm SFR}$ values by using our $R_{\rm opt}/R_{\rm radio}$ ratios, they will increase by a factor of $\sim0.7\,(0.4)$\,dex  at redshift 1.5 (2), which would alleviate this  observed discrepancy. Lastly, in Fig. \ref{fig:surfacedensity} we present the sample of SFGs from \cite{elbaz18} and Lang et al. (in prep.). Although  $z\sim1.5$ SFGs also follow a positive relation between $\log(\Sigma_{\rm SFR})$ and $\Delta\log(\rm SSFR)_{\rm MS}$, those at $z\sim2$  are widely scattered. As reported by \citet{elbaz18},  $z\sim2$ SFGs with starburst-like $\Sigma_{\rm SFR}$ are also found close to/within the MS. According to our results, there is a fraction of MS galaxies for which $\Sigma_{\rm SFR}$ is significantly higher  than expected from  the  $\log(\Sigma_{\rm SFR})-\Delta\log(\rm SSFR)_{\rm MS}$ relation. These 
high-$\Sigma_{\rm SFR}$ MS galaxies  are present at all redshifts  and comprise less than $10\%$ of the MS galaxy population (see contour levels of Fig. \ref{fig:surfacedensity}).

\begin{figure}
	\centering
	\includegraphics[width=9.2cm]{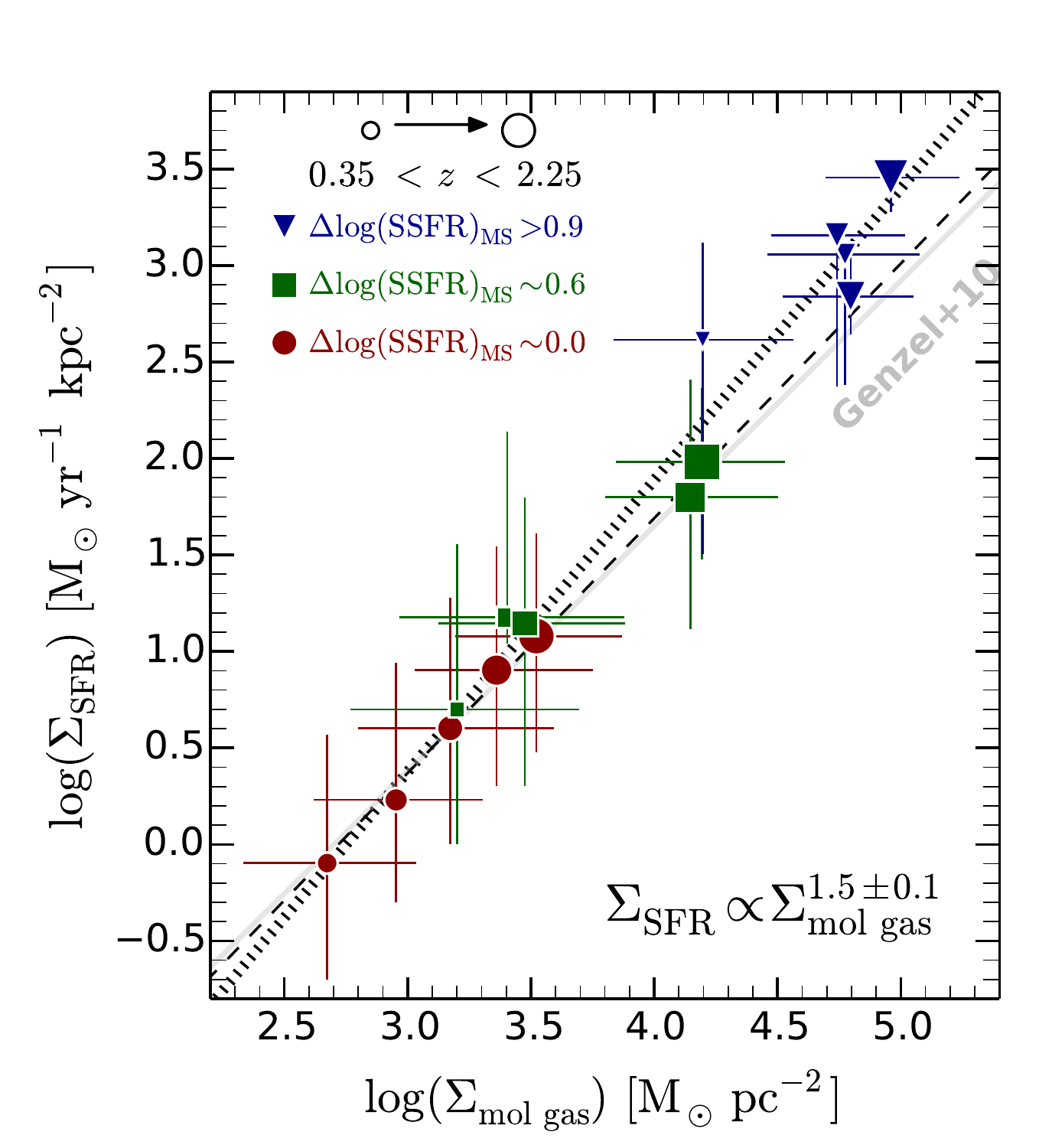}
	\caption{SFR surface density ($\Sigma_{\rm SFR}$) as a function of the molecular gas surface density ($\Sigma_{\rm mol\,gas}$). The data points show the locus of the median $\Sigma_{\rm SFR}$ and $\Sigma_{\rm mol\,gas}$ of galaxies binned in $\Delta\log(\rm SSFR)_{\rm MS}$ and redshift -- as in Fig. \ref{fig:surfacedensity}. The symbol size increases with redshift, while the color indicates the median $\Delta\log(\rm SSFR)_{\rm MS}$. The solid gray line shows the KS relation reported by \citet{genzel10}, adapted for a Salpeter IMF. While the dotted black line illustrates the best linear fit to all the data points, the dashed thin line shows the best linear fit when we exclude SFGs with $\Delta\log(\rm SSFR)_{\rm MS}>0.9$.  Error bars represent the 16th and 84th percentile of the inferred $\Sigma_{\rm SFR}$ and $\Sigma_{\rm mol\,gas}$ distributions. The molecular gas mass has been approximated  by using the prescription of  \citet[][see Table 4]{genzel15};  where  $M_{\rm mol\,gas}=M_{\rm mol\,gas}(z, {\rm SSFR}, M_\star=10^{10.5\pm0.5}\rm M_\odot)$. }
	\label{fig:ksrelation}%
\end{figure}


\begin{figure*}[h!]
	\centering
	\includegraphics[width=18.5cm]{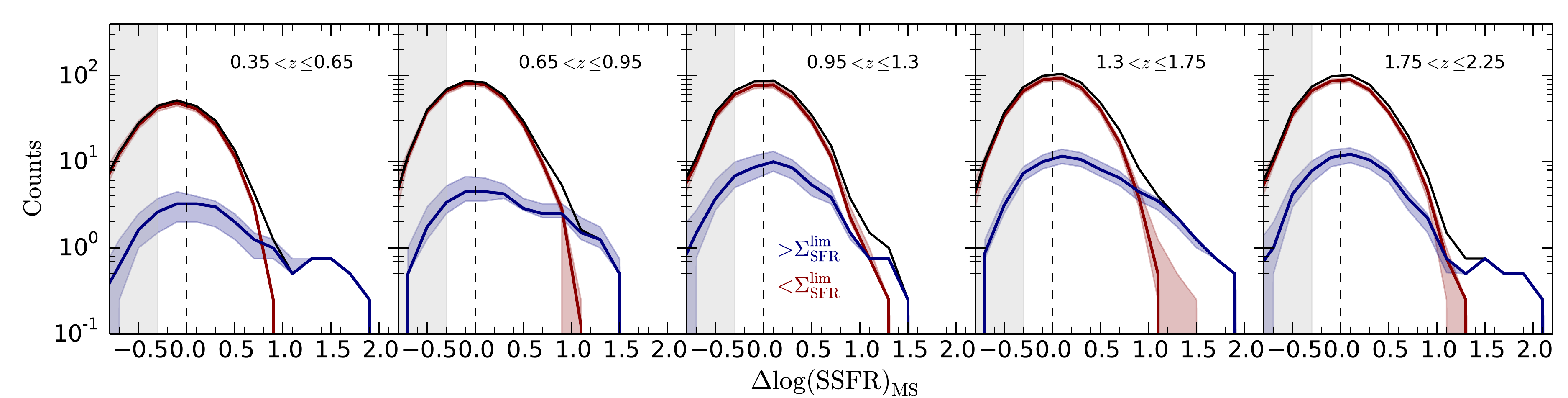}
	\caption{Distribution of SFGs along $\Delta\log(\rm SSFR)_{MS}$ (black solid line). In this illustrative case, we separate the  main-sequence (red line) and starburst (blue) contribution using $\Sigma_{\rm SFR}^{\rm lim}\equiv \Sigma_{\rm SFR}(\Delta\log(\rm SSFR)_{\rm MS}=0.7)$ for SFGs with $\log(M_\star/{\rm M_\odot})> 10.5$. The red and blue shaded regions show the scatter (16th and 84th percentile) introduced by the uncertainties and upper limits of our measurements. The gray region indicates the parameter space where our sample is not complete. }
	\label{fig:doublegaussian}%
\end{figure*}

\section{Discussion}\label{sec:discussion}

\subsection{The cold gas accretion {\it vs} merger mode of star formation  }\label{subsubsec:discussion}
 We have revealed that most SFGs ($\log(M_\star/\rm M_\odot)>10.5$) follow a linear relation in the $\log(\Sigma_{\rm SFR})-\Delta\log(\rm SSFR)_{\rm MS}$ plane over the redshift range $0.35<z<2.25$. 
 To first order,   these results can be  discussed within the context of the Kennicutt-Schmidt (KS) relation \citep[$\Sigma_{\rm SFR}-\Sigma_{\rm gas}$;][]{kennicutt98}. We therefore use the scaling relations of \citet{genzel15} to derive the typical molecular gas mass of galaxies at three different $\Delta\log(\rm SSFR)_{MS}$ bins, namely  [$-0.3$,0.3], [0.3,0.6] and [0.9, 2.0] as in Fig. \ref{fig:surfacedensity}. Then, we assume that our radio size estimates (Table \ref{table:2})  also trace the extent of the molecular gas reservoir; these are subsequently used to approximate the galactic averaged molecular gas density \hbox{$(\Sigma_{\rm mol\,gas}=M_{\rm mol\,gas}/2\pi R_{\rm eff})$.} This information is combined with the $\Sigma_{\rm SFR}$ values presented in Table \ref{table:3}, allowing us to approximate the shape of the KS relation (see Fig. \ref{fig:ksrelation}). 
  It is reassuring that our data points, covering a wide range in redshift and $\Delta\log(\rm SSFR)_{\rm MS}$, agree within the uncertainties with the KS relation presented by \citet{genzel10}. Moreover,  our data points are consistent with the scenario wherein low and high-redshift SFGs follow a similar molecular gas–star formation relation \citep{bouche07, genzel10}. By considering  SFGs with $-0.3 \lesssim \Delta\log(\rm SSFR)_{\rm MS}\lesssim 0.9$, we derive a super-linear slope of  $1.3\pm0.1$. If SFGs with  $\Delta\log(\rm SSFR)_{\rm MS}>0.9$ are included, the slope becomes steeper, i.e. $1.5\pm0.1$; which  indicates that SFGs evolve towards a more efficient regime of star formation as $\Delta\log(\rm SSFR)_{\rm MS}$ increases.   This is consistent with the small size, and hence higher $\Sigma_{\rm SFR}$, of galaxies above the MS (Fig. \ref{fig:size-dms}); which could be the result of gas-rich mergers  \citep[e.g., ][]{moreno15, wellons15} and/or violent disk instability \citep[VDI; e.g., ][]{dekelburkert14, tacchella16, wang18}. 

Beyond the broad picture of galaxy evolution discussed above, we have also reported the discovery of  MS galaxies harboring starburst-like $\Sigma_{\rm SFR}$ conditions (Fig. \ref{fig:surfacedensity}). This result echoes, in particular, that of \cite{elbaz18},  who reported the presence of ``hidden'' starbursts within the MS at $z\sim2$. Then, the fundamental question  arises as: what is the physical mechanism responsible for high-$\Sigma_{\rm SFR}$ MS galaxies? Firstly, these systems could  be a result of  large cold gas reservoirs distributed over small disk radii \citep{wang18}  that, due to disk instability episodes \citep[e.g., ][]{dekelburkert14}, yield high $\Sigma_{\rm SFR}$.  If the gas replenishment occurs when a galaxy lies at the lower envelope of the MS (i.e., \hbox{$\Delta\log(\rm SSFR)_{\rm MS}\sim-0.3$}), the SFR enhancement might not suffice to bring the galaxy above the MS.  Secondly, high-$\Sigma_{\rm SFR}$ MS galaxies could be explained in the context of merger-driven bursts of star formation; depending on the gas content,  mergers  could not  significantly increase the SFR and offset the galaxy from the MS.  \citep[e.g., ][]{fensch17, wang18}.  This is in agreement with the observational evidence of merging activity in galaxies on and above the MS   \citep[e.g., ][Karteltepe priv. communication]{karteltepe12, ellison18, calabro18, wang18, cibinel18}. \\

\begin{figure}
	\centering
	\includegraphics[width=8.5cm]{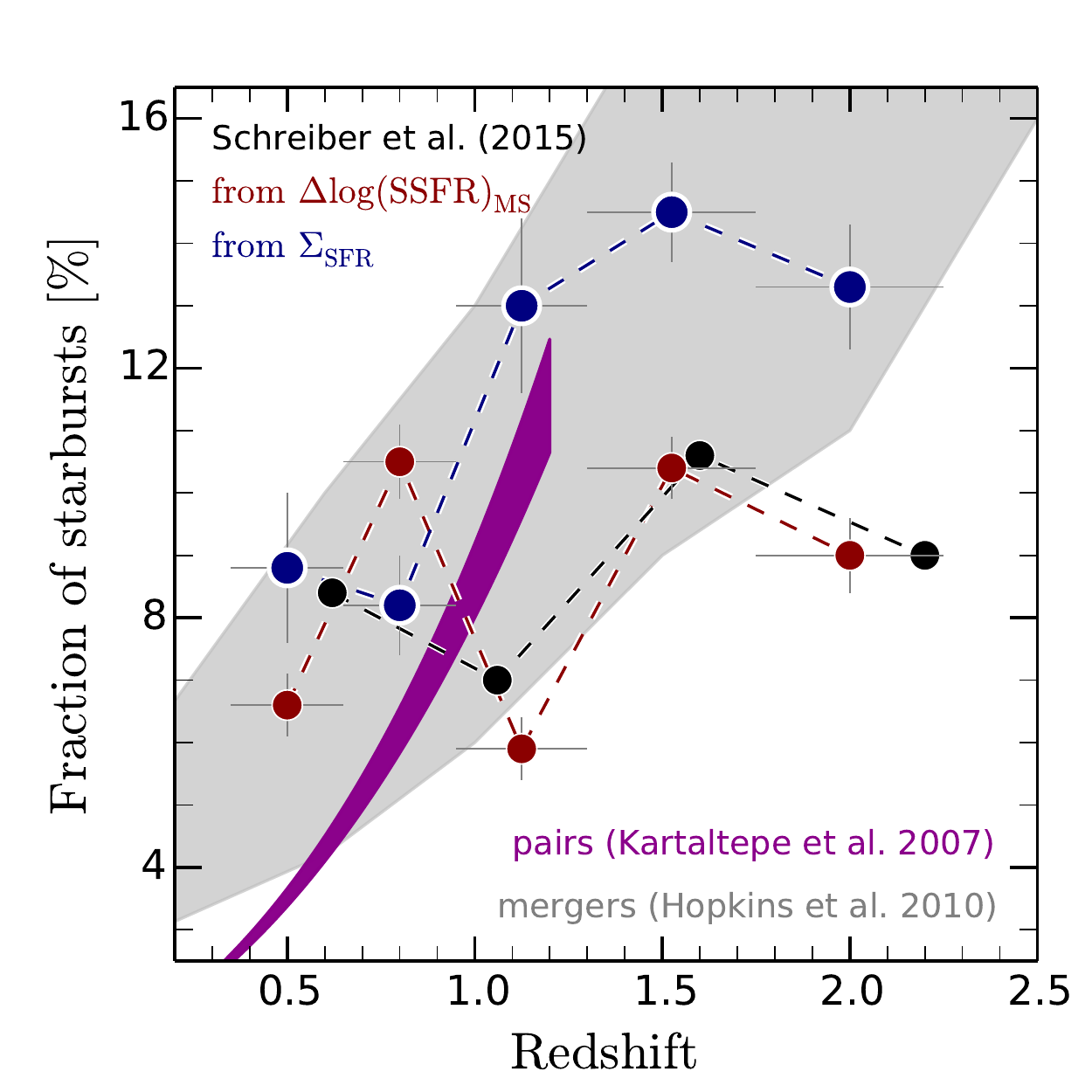}
	\caption{ Redshift evolution of the ``starburst" fraction from the  mass-complete sample of radio-selected SFGs used in this work. We adopt two definitions of a starburst galaxy: a)  systems with $\Delta\log(\rm SSFR)>0.39$ and b) $\Sigma_{\rm SFR}> \Sigma_{\rm SFR}^{\rm lim}$  (see Sect. \ref{subsubsec:discussion}), in both cases  $\log(M_\star/\rm M_\odot)>10.5$.  For comparison, we present the ``starburst" fraction for $\log(M_\star/\rm M_\odot)>10.5$ and $\Delta\log(\rm SSFR)>0.39$ UV/FIR-selected SFGs from \citet{schreiber15}. The redshift evolution of the observed  galaxy pair fraction  is given by the magenta shaded region \citep{karteltepe07}. Grey region shows the major merger fraction predicted by \cite{hopkins10}. }
	\label{fig:frac_redshift}%
\end{figure}

 In light of these findings, {$\Sigma_{\rm SFR}$ arises as a remarkable proxy to identify starburst galaxies, where star formation is triggered by either mergers or  VDI that lead to high  $\Sigma_{\rm SFR}$.} 
 As an illustrative case, we here  evaluate  $\Sigma_{\rm SFR}^{\rm lim}\equiv Mdn[\Sigma_{\rm SFR}(\Delta\log(\rm SSFR)_{\rm MS}=0.7)]$\footnote{We use this threshold as  the number of $z\sim0.35$ starbursts is consistent with that derived from the standard $\Delta\log(\rm SSFR)_{MS}$--based  definition (see Fig. \ref{fig:frac_redshift}). We note that a different threshold in $\Sigma_{\rm SFR}$ also yields a larger starburst fraction at high redshift. }, at each redshift bin (Sect. \ref{subsec:surfacedensity}), and adopt it as a threshold to identify starbursting systems. Under this definition,  it is possible to  decompose the bimodal distribution of SFGs along $\Delta\log(\rm SSFR)_{MS}$ \citep[e.g., ][]{sargent12} into main-sequence ($<\Sigma_{\rm SFR}^{\rm lim}$) and starburst ($>\Sigma_{\rm SFR}^{\rm lim}$) contribution (see Fig. \ref{fig:doublegaussian}). The first, and more dominant,  distribution is centered at $\Delta\log(\rm SSFR)_{MS}=0$ and represents  the population of galaxies forming stars through a secular mode of star formation  \citep[e.g., ][]{dekel09,  sellwood14}. The distribution of non-merger and merger-induced starbursts exhibits an enhanced tail at high $\Delta\log(\rm SSFR)_{MS}$ and, consequently, its median lies at $\Delta\log(\rm SSFR)_{MS}>0$.  
 We note that this $\Sigma_{\rm SFR}^{\rm lim}$--based  scheme brings the galaxy-pair/merger rate in a better agreement with the fraction of high-redshift starbursts  (see Fig. \ref{fig:frac_redshift}), given that a   $\Delta\log(\rm SSFR)_{MS}$--based definition  misses the merger-induced starbursts ``hidden'' within the MS \citep{elbaz18, cibinel18}.

\subsection{Centrally-concentrated star formation; evidence of bulge growth?}\label{subsec:discussion_bulge}
The finding of compact radio continuum emission of SFGs on and above the MS  further support the evidence of star formation enhancement at small radii \cite[e.g., ][]{simpson15, rujopakarn16, nelson16, fujimoto17, elbaz18}. Interestingly,  while the extent of the stellar component  increases with cosmic time, the overall region where most stars are formed  remains nearly constant (see Fig. \ref{fig:size-redshift}). This might indicate that  fresh star-forming gas is constantly fueled towards the center of galaxies, either due to VDI,  minor/major mergers and/or tidal interactions \citep[e.g., ][]{larson03, rupke10, sillero17, ellison18, munoz-elgueta18}. Regardless of the dominant mechanism driving the formation of stars in galaxies (on and above the MS),  violent and secular galaxy's evolutionary channels  lead to the formation of a bulge \citep[e.g., ][]{kormendy04, fisher06, hopkins09,  brooks16,  tonini16}. Ultimately, the presence of a dominant bulge could  stabilize the gas disk against gravitational instabilities and hence prevent the formation of stars \citep[e.g., ][and references therein]{lang14}.

In this context,  we hypothesize that the centrally concentrated star formation activity of most SFGs  in the sample might reflect the growth of the bulge, which might precede the quenching of the galaxy from the inside-out \citep[e.g., ][]{ellison18}. At this late evolutionary stage, the bulge of massive galaxies is fully quenched, while star formation activity still take place at large radius \citep[e.g., ][]{tacchella15, rowlands18}.  Spatially resolved studies of low and high-mass SFGs at high redshift are needed to verify such a scenario, allowing us to understand how star formation, and hence stellar mass, is distributed in galaxies across cosmic time.

\section{Summary}\label{sec:summary}
We  presented the first systematic study of the radio continuum size evolution of SFGs over $0.35<z<2.25$.  We used a mass-complete sample of 3184  radio-selected SFGs, detected in the  VLA COSMOS 3GHz map \citep{smolcic17}, and performed extensive Monte Carlo simulations to characterize our selection function and validate the robustness of our  measurements.  We  found that:  

\begin{itemize}
	\item The radio continuum size shows no clear dependence on the stellar mass  of SFGs with $10.5\lesssim \log(M_\star/ \rm M_\odot)\lesssim11.5$, that is  the mass range where our sample is not affected by our selection function. \\

	\item MS galaxies are preferentially  (but not exclusively)  extended, while SFGs above the MS are more compact systems; the median size of SFGs on (above) the MS is $R_{\rm eff}=1.5\pm0.2\,(1.0\pm0.2)$\,kpc. Using the parametrization of the form $R_{\rm eff}\propto (1+z)^\alpha$, we found that the median size remains nearly constant with cosmic time, with $\alpha=-0.26\pm0.08\,(0.12\pm0.14)$ for SFGs on (above) the MS.\\
	
	\item The median radio size of SFGs  is smaller (by a factor $1.3-2$) than that inferred from optical-to-UV emission that traces their stellar component \citep{shibuya15}. 	Overall, these results are consistent with compact radio continuum, FIR and extinction-corrected H$\alpha$ emission \citep[$\lesssim1.5$\,kpc; e.g., ][]{nelson16, rujopakarn16, murphy17, cotton18, elbaz18, lindroos18}.  \\

	\item Most SFGs follow a linear relation in the $\log(\Sigma_{\rm SFR})-\Delta\log(\rm SSFR)_{MS}$ plane, consistent with previous studies of SFGs in the local Universe \citep{lutz16} and at $z\sim1$ \citep{magdis16}. While  its normalization increases with redshift, its slope becomes steeper at lower redshifts (from $\alpha=1.5$ at $z\sim2$ to 2.6 at $z\sim0.5$). \\
	
	\item There is a fraction ($\lesssim10\%$) of MS galaxies harboring starburst-like $\Sigma_{\rm SFR}$, consistent with recent evidence of  ``hidden'' starburst within the MS  at $z\sim2$ \citep{elbaz18}.
	
\end{itemize}

Overall, our results suggest that SFGs with enhanced star formation undergo a compaction phase. These systems could be explained in the context of disk instability  and/or merger-driven burst of star formation that, depending on the gas content, offset the galaxy from the MS in different proportions \citep[e.g., ][]{fensch17,wang18}. Since using $\Delta\log(\rm SSFR)$ alone prevents us from identifying those starburst galaxies \hbox{"hidden"} within the MS, we recommend to use in addition $\Sigma_{\rm SFR}$ to better identify starbursting systems. Having constraints on $\Sigma_{\rm SFR}$ is the first step towards the characterization of the KS relation at high redshift. Exploring  in detail the gas content, and optical morphology, of SFGs in our sample is the subject of an upcoming manuscript.

\begin{acknowledgements}
E.F.J.A,  B.M., A.K, F.B., E.V., E.R.D., K.H. and C.K.  acknowledge  support of the Collaborative Research Center 956, subproject A1, funded by the Deutsche Forschungsgemeinschaft (DFG).  E.F.J.A. thanks A. Mikler and J. Erler for  their encouragement and insightful comments. E.V. acknowledges funding from the DFG grant BE 1837/13-1. P.L. acknowledges funding from the European Research Council (ERC) under the European Union’s Horizon 2020 research and innovation programme (grant agreement No. 694343). S.T. acknowledges support from the ERC Consolidator Grant funding scheme (project ConTExt, grant No. 648179). The Cosmic Dawn Center is funded by the Danish National Research Foundation. J.D. acknowledges the financial assistance of the South African Radio Astronomy Observatory (SARAO; www.ska.ac.za). Based on observations with the National Radio Astronomy Observatory which is a facility of the National Science Foundation operated under cooperative agreement by Associated Universities, Inc. Based also on data products from observations made with ESO Telescopes at the La
Silla Paranal Observatory under ESO programme ID 179.A-2005 and on
data products produced by TERAPIX and the Cambridge Astronomy Survey
Unit on behalf of the UltraVISTA consortium. 
\end{acknowledgements}

%
%

\bibliographystyle{aa} 
\bibliography{jimenez-andrade+19.bib} 

\begin{appendix}
\twocolumn

\section{Flux boosting}\label{appendix:c}

\begin{figure}[h!]
	\centering
	\includegraphics[width=8.cm]{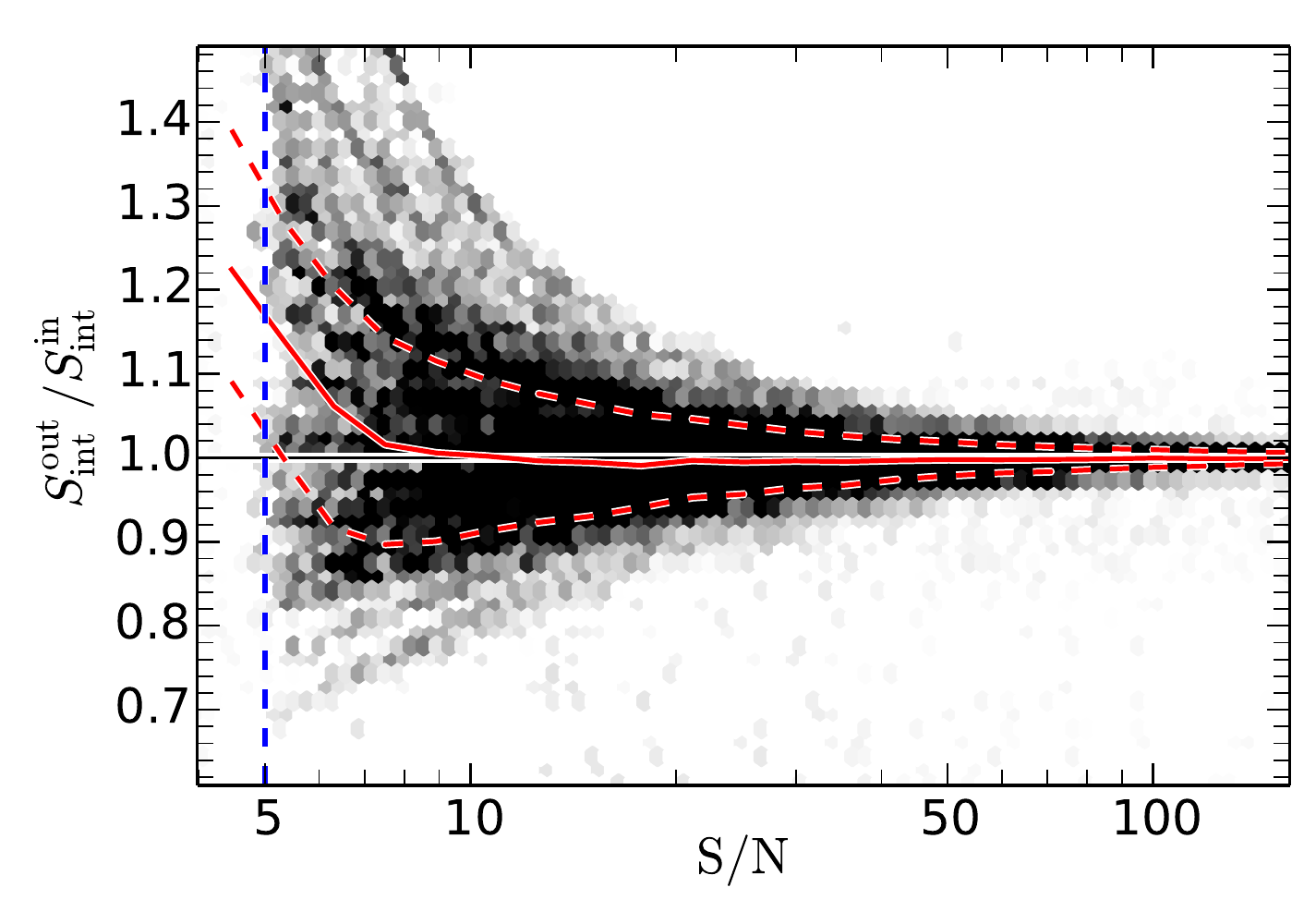}
	\caption{Flux boosting for unresolved sources as a function of S/N estimated from MC simulations.  The solid and dashed red lines show the 50th percentile, 14th and 84th percentile of the distribution as a function of S/N.  The vertical blue dashed line indicates our detection threshold (S/N=5).  }
	\label{fig:fluxboosting_unresolved}
\end{figure}

\begin{figure}[h]
	\centering
	\includegraphics[width=9.2cm]{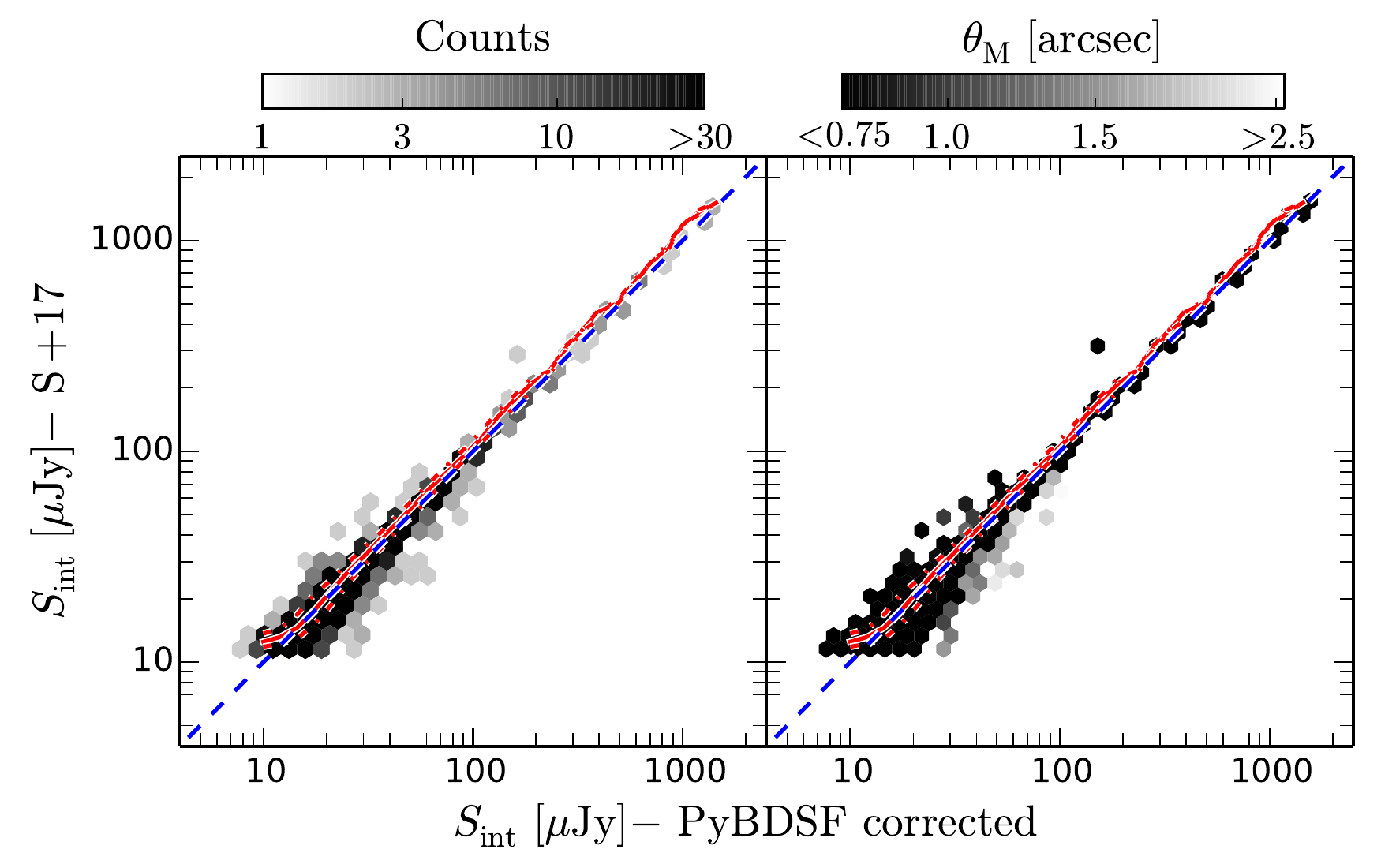}
	\caption{Comparison between the flux density of 3184 SFGs in the sample (resolved and unresolved) derived from {\tt PyBDSF} (corrected) and those reported by \citet{smolcic17}. The solid and dashed red lines show the 50th percentile, 14th and 84th percentile of the distribution  as a function of the flux density reported in this study. }
	\label{fig:blobcatvspybdsf}%
\end{figure}

\newpage
\onecolumn
\section{Size evolution of SFGs using non-corrected FWHM and flux density}\label{appendix:b}

\begin{figure*}[h!]
	\centering
	\includegraphics[width=18.2cm]{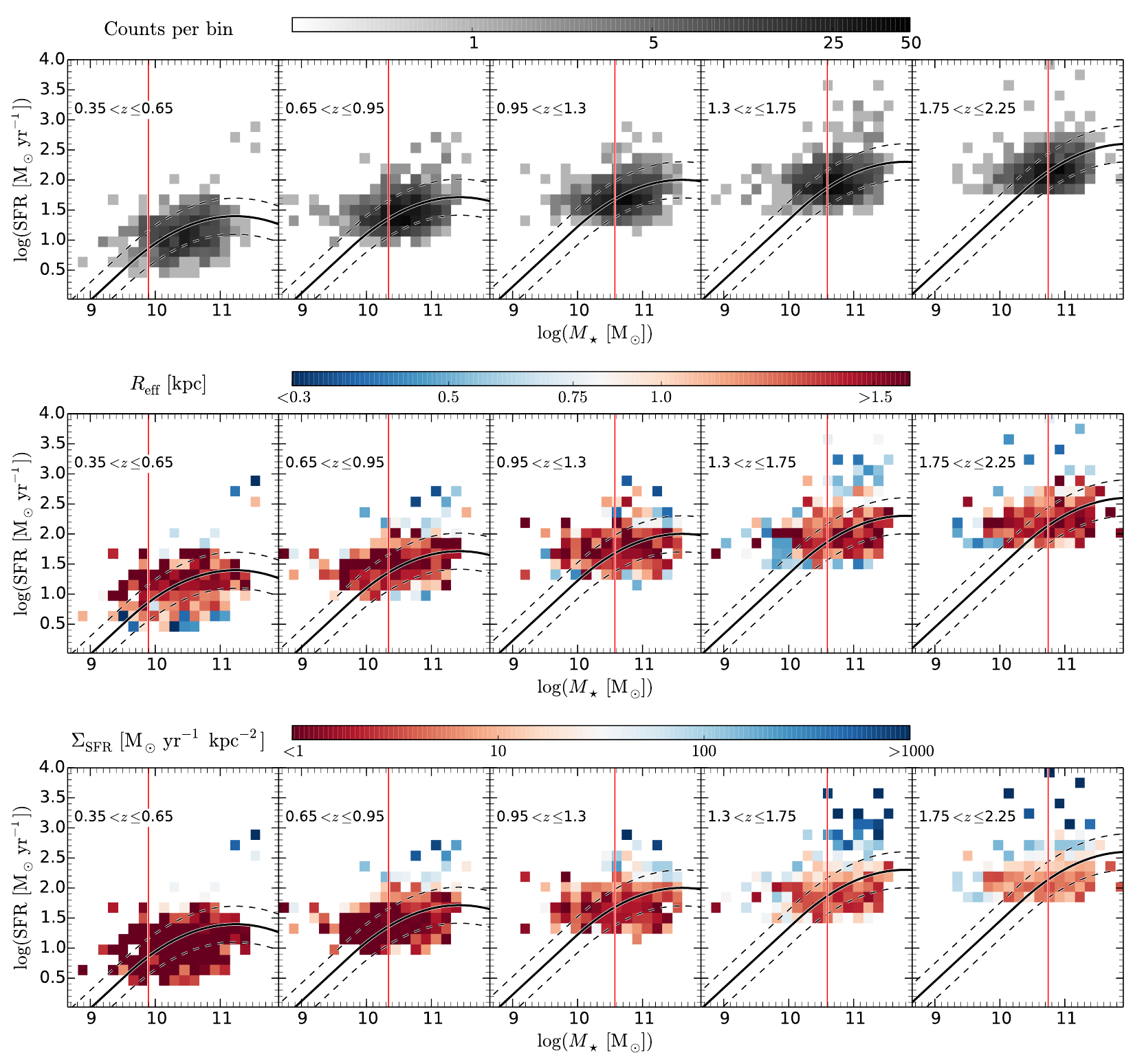}
	\caption{{\it Upper panels:} 3184 SFGs in the ${\rm SFR}-M_{\star}-z$ plane.  The color scale shows the number of sources per bin. Solid/dashed black lines show the position of the MS of SFGs and its associated dispersion given by \citet{schreiber15}. Vertical red solid lines show the mass-limit above which we consistently probe SFGs on and above the MS. {\it Middle and lower  panels:} median size  and star formation surface density, respectively, of 3184 SFGs in the sample. We use 1000 MC realizations to include the values for unresolved sources, which are drawn from the distributions described in Sects. \ref{subsec:size-dms}   and  \ref{subsec:surfacedensity}. } 
	\label{fig:a_sfr-mass-z_plane}%
\end{figure*}

\begin{figure*}[h!]
	\centering
	\includegraphics[width=18.0cm]{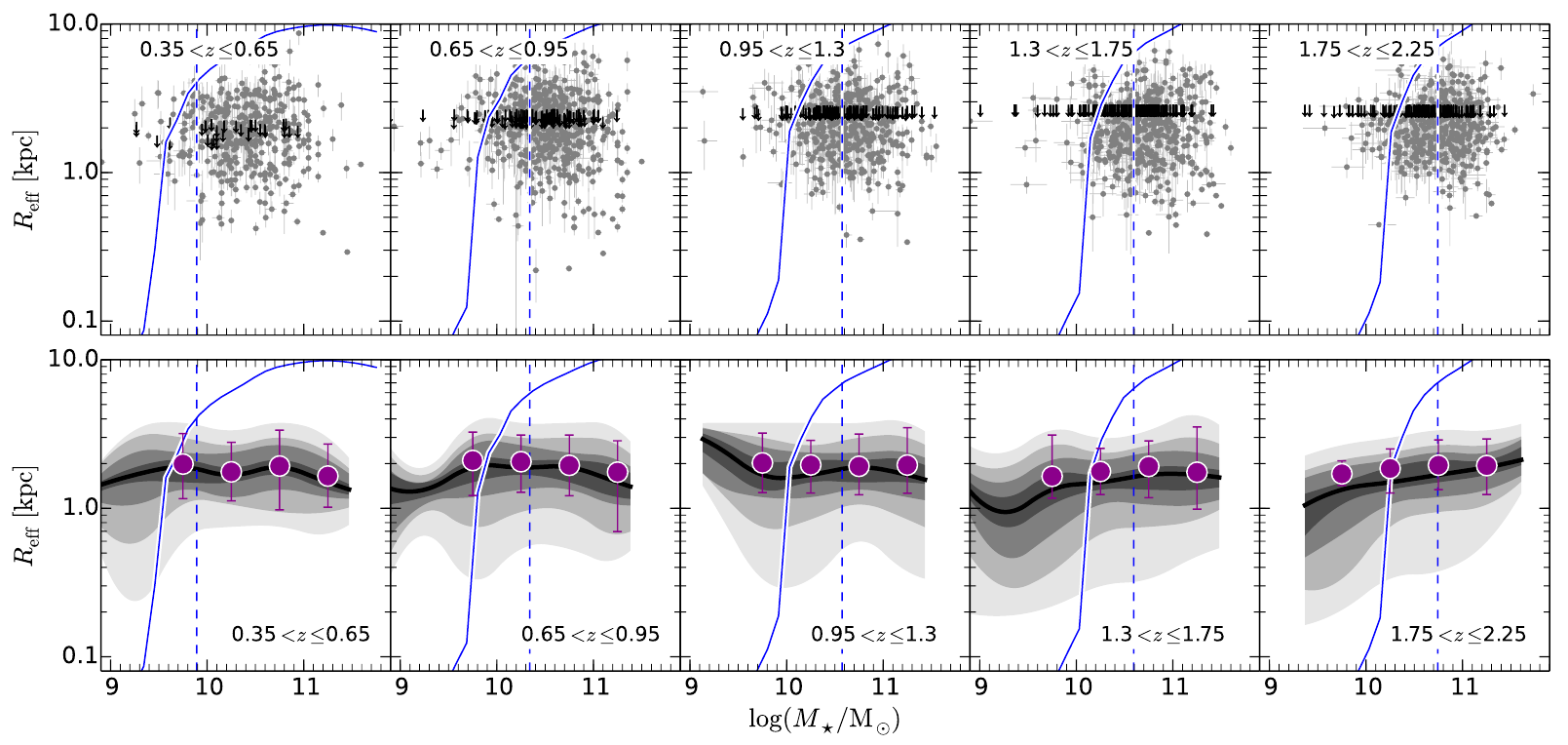}   
	\caption{3184 SFGs in the $R_{\rm eff}-M_{\star}-z$ plane.  	The blue solid line shows the  maximum size, corresponding to the 10\% completeness level,  that can be observed for a galaxy with $\Delta \log(\rm SSFR)=0$  evaluated at the central redshift value per bin. 	Blue dashed lines show the mass-limit above which we consistently probe SFGs on and above the MS. Black arrows show the upper limits for the size of unresolved sources. {\it Lower panels}: density distribution per stellar mass bin (0.5 dex width) of SFGs in the  $R_{\rm eff}-M_{\star}-z$ plane. Contour levels are at the 10, 20, 30, 40, 50, 60, 70, 80 and 90th  percentile. The median size derived via the KM estimator is shown by the dark magenta circles, the error bars correspond to the 16th and 86th percentile of the distribution.  }
	\label{fig:a_mass-size_plane}%
\end{figure*}

\begin{figure*}[h!]
	\centering
	\includegraphics[width=18.0cm]{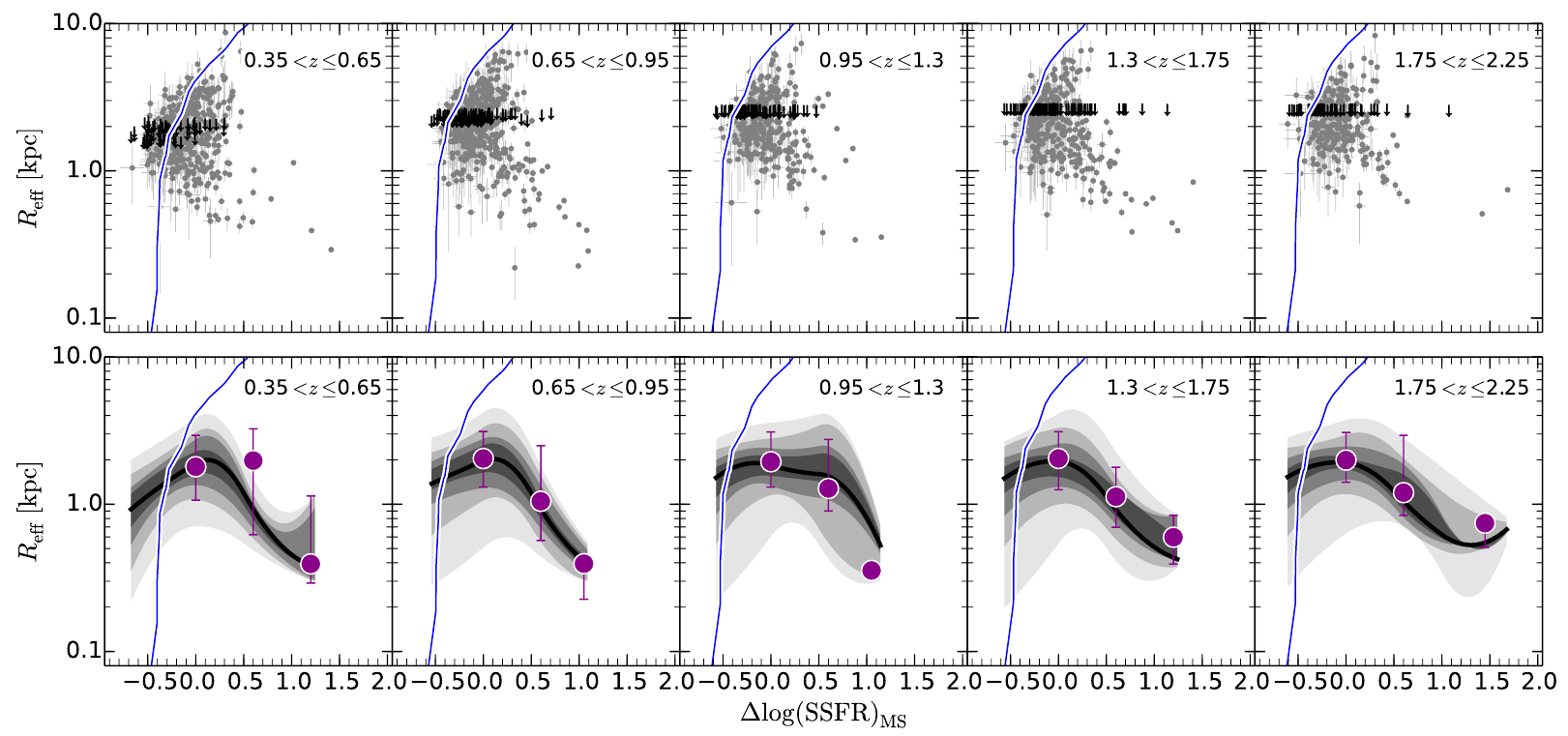}    
	\caption{{\it Upper panels:}  SFGs with $M_\star>M_\star^{\rm lim}$  in the size$-\Delta\log({\rm SSFR})-z$ plane.  	The blue solid line shows the  maximum size, corresponding to the 10\% completeness level,  that can be observed for a galaxy with $M_\star=M_\star^{\rm lim}$ evaluated at the central redshift value per bin. 	Black arrows show the upper limits for the size of unresolved sources.   {\it Lower panels:} density distribution per $\Delta\log(\rm SSFR)$ bin (0.5 dex width) of SFGs in the  size$-\Delta\log({\rm SSFR})-z$ plane. Contour levels are at the 10, 20, 30, 40, 50, 60, 70, 80 and 90th  percentile. The median size derived via the KM estimator is shown by the dark magenta circles, the error bars correspond to the 16th and 86th percentile of the distribution.  } 
	\label{fig:a_ssfr-size_plane}%
\end{figure*}

\begin{figure*}[h!]
	\centering
	\includegraphics[width=18cm]{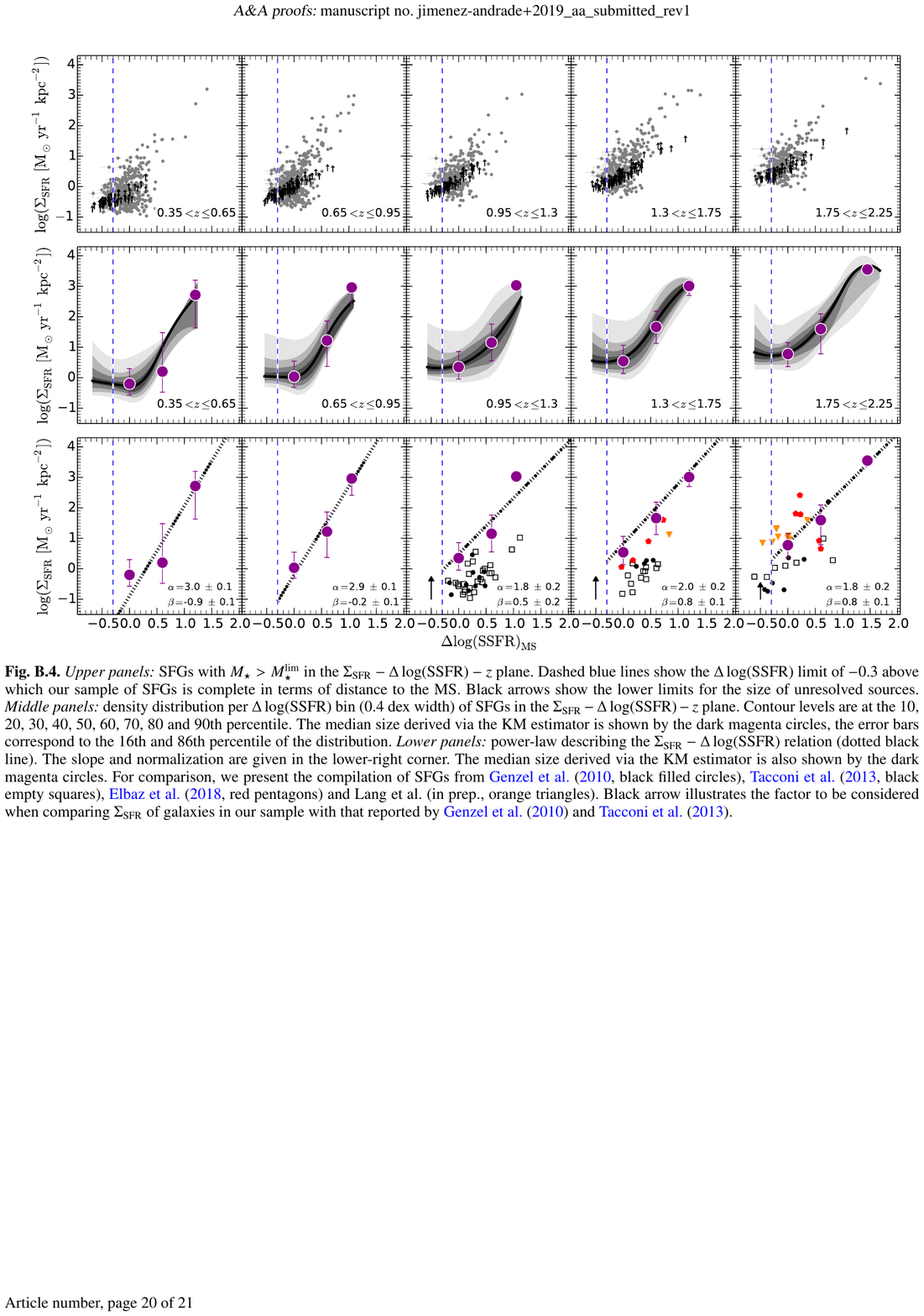}
	\caption{{\it Upper panels:} SFGs with $M_\star>M_\star^{\rm lim}$  in the  $\Sigma_{\rm SFR}-\Delta\log({\rm SSFR})-z$ plane. 	Dashed blue lines show the $\Delta\log(\rm SSFR)$ limit of $-0.3$ above which our sample of SFGs is complete in terms of distance to the MS. 	Black arrows show the lower limits for the size of unresolved sources.   {\it Middle panels:} density distribution per $\Delta\log(\rm SSFR)$ bin (0.4 dex width) of SFGs in the  $\Sigma_{\rm SFR}-\Delta\log({\rm SSFR})-z$ plane. Contour levels are at the 10, 20, 30, 40, 50, 60, 70, 80 and 90th  percentile. The median size derived via the KM estimator is shown by the dark magenta circles, the error bars correspond to the 16th and 86th percentile of the distribution.   {\it Lower panels:} power-law describing the $\Sigma_{\rm SFR}-\Delta\log(\rm SSFR)$ relation (dotted black line). The slope and normalization are given in the lower-right corner. The median size derived via the KM estimator is also shown by the dark magenta circles.  For comparison, we present the compilation of SFGs from \citet[][black filled circles]{genzel10}, \citet[][black empty squares]{tacconi13},   \citet[][red pentagons]{elbaz18} and Lang et al. (in prep., orange triangles).  Black arrow illustrates the factor to be considered when comparing  $\Sigma_{\rm SFR}$ of galaxies in our sample with that reported by \cite{genzel10} and \cite{tacconi13}.   }
	\label{fig:a_ssfr-dms-plane}%
\end{figure*}

The extensive Monte Carlo simulations performed in this work indicate that the FWHM  and flux density are being overestimated for most of the radio sources in the sample (see Fig. \ref{fig:comparison}). 
 Using these values, however, does not systematically affect the trends and relations presented in this work -- as initially inferred from Fig. \ref{fig:a_sfr-mass-z_plane}. First, this approach also leads to a flat relation between the  median radio size and stellar mass of SFGs (Fig. \ref{fig:a_mass-size_plane}). Second, we  find a similar radio size/$\Sigma_{\rm SFR}$ dichotomy between SFGs on and above the MS  (Fig. \ref{fig:a_sfr-mass-z_plane} and  \ref{fig:a_ssfr-size_plane}). Using non-corrected measurements does lead to a smaller fraction of MS galaxies with starburst-like $\Sigma_{\rm SFR}$ (Fig. \ref{fig:a_ssfr-dms-plane}), given that the size of faint MS galaxies is overestimated and, consequently,  $\Sigma_{\rm SFR}$ becomes smaller. We note that regardless the use of corrected or non-corrected values, the fraction of  starburst-like $\Sigma_{\rm SFR}$ systems  remains unclear due to the presence of MS galaxies that are unresolved in the VLA COSMOS 3GHz map.

\onecolumn
\section{Tables}

\begin{table*}[h!]
	\begin{center}
		\caption{Radio continuum size  as a function of the stellar mass of SFGs}
		{\small
			\begin{tabular}{c | c c c c c c }
				\hline 
				\hline  \\[-4pt] 
				$\log(M_\star/\rm M_\odot)$ [dex]  &   $0.35<z<0.65$ &   $0.65<z<0.95$ &  $0.95<z<1.30$  &   $1.30<z<1.75$ &   $1.75<z<2.25$   \\[2pt] 
				\hline					 \\[-5pt] 
				$[10.0, 10.5]$  &   $1.5^{+1.2}_{-0.8}$ &   $1.5^{+1.4}_{-0.9}$ &   \dots\tablefootmark{a}  &    \dots &   \dots  \\[5pt]		
				$[10.5, 11.0]$   &   $1.8^{+1.6}_{-1.1}$ &   $1.5^{+1.4}_{-0.7}$ &  $1.4^{+1.6}_{-0.7}$  &   $1.3^{+1.2}_{-0.8}$ &   $1.2^{+1.4}_{-0.5}$ \\[5pt]	
				$[11.0, 11.5]$   &   $1.5^{+1.4}_{-0.6}$ &   $1.3^{+1.4}_{-0.8}$ &  $1.7^{+1.4}_{-0.9}$  &   $1.2^{+2.2}_{-0.5}$ &   $1.5^{+1.2}_{-0.8}$ \\[5pt]		
				\hline 
		\end{tabular} }
		\tablefoot{The effective radius is given in kpc. The uncertainties correspond to the 16th and 84th percentile of the size distribution per stellar mass  bin.  \tablefootmark{a} No vualues are given for the mass bins that are strongly affected by incompleteness.  }
		\label{table:1}	
	\end{center} 
\end{table*}

\begin{table*}[h!]
	\begin{center}
		\caption{Radio continuum size  as a function of distance to the MS of SFGs}
		{\small
			\begin{tabular}{c | c c c c c c }
				\hline 
				\hline  \\[-4pt] 
				$\Delta\log(\rm SSFR)_{\rm MS}$ [dex]   &   $0.35<z<0.65$ &   $0.65<z<0.95$ &  $0.95<z<1.30$  &   $1.30<z<1.75$ &   $1.75<z<2.25$   \\[2pt] 
				\hline					 \\[-5pt] 
				$[-0.3, 0.3]$  &   $1.7^{+1.3}_{-1.0}$ &   $1.5^{+1.3}_{-0.8}$ &   $1.4^{+1.5}_{-0.8}$  &    $1.5^{+1.2}_{-0.9}$ &   $1.5^{+1.2}_{-0.7}$  \\[5pt]		
				$[0.3, 0.9]$   &   $1.0^{+2.0}_{-0.7}$ &   $1.0^{+2.0}_{-0.6}$ &  $1.3^{+1.7}_{-0.7}$  &   $0.9^{+0.6}_{-0.5}$ &   $0.9^{+1.0}_{-0.3}$ \\[5pt]	
				$>\,0.9\tablefootmark{a}$   &   $0.4^{+0.8}_{-0.1}$ &   $0.4^{+0.1}_{-0.2}$ &  $0.5^{+0.1}_{-0.2}$  &   $0.7^{+0.2}_{-0.2}$ &   $0.8^{+0.1}_{-0.3}$ \\[5pt]		
				\hline 
		\end{tabular} }
		\tablefoot{The effective radius is given in kpc. The uncertainties correspond to the 16th and 84th percentile of the size distribution per $\Delta\log(\rm SSFR)_{\rm MS}$  bin. The minimum stellar mass  probed thoughtout the different redshift bins is  $\log(M_\star^{\rm lim}/\rm M_\odot)=9.9, 10.2, 10.5, 10.5, \, \rm and \, 10.7$, respectively. \tablefoottext{a}{The highest $\Delta\log(\rm SSFR)_{\rm MS}$  bin is centered at $\Delta\log(\rm SSFR)_{\rm MS}=1.2, 1.05, 1.05, 1.45, 1.45$, respectively. } 
		}
		\label{table:2}	
	\end{center} 
\end{table*}

\begin{table*}[h!]
	\begin{center}
		\caption{Star formation surface density $(\Sigma_{\rm SFR})$ as a function of distance to the MS of SFGs}
		{\small
			\begin{tabular}{c | c c c c c c }
				\hline 
				\hline  \\[-4pt] 
				$\Delta\log(\rm SSFR)_{\rm MS}$ [dex]  &   $0.35<z<0.65$ &   $0.65<z<0.95$ &  $0.95<z<1.30$  &   $1.30<z<1.75$ &   $1.75<z<2.25$   \\[2pt] 
				\hline					 \\[-5pt] 
				$[-0.3, 0.3]$  &   $0.8^{+2.9}_{-0.6}$ &   $1.7^{+7.0}_{-1.2}$ &  $4^{+15}_{-3}$  &   $8^{+27}_{-6}$ &   $12^{+29}_{-9}$   \\[5pt]		
				$[0.3, 0.9]$   &   $5^{+31}_{-4}$ &   $15^{+123}_{-14}$ &  $14^{+49}_{-12}$  &   $63^{+194}_{-50}$ &   $96^{+135}_{-66}$ \\[5pt]	
				$>\,0.9\tablefootmark{a}$    &   $412^{+900}_{-380}$ &   $1140^{+190}_{-900}$ &  $1435^{+100}_{-1200}$  &   $690^{+610}_{-250}$ &   $2860^{+100}_{-970}$ \\[5pt]		
				\hline 
		\end{tabular} }
		\tablefoot{$\Sigma_{\rm SFR}$ is given in $\rm M_\odot\,yr^{-1} \, kpc^{-2}$. The uncertainties correspond to the 16th and 84th percentile of the $\Sigma_{\rm SFR}$ distribution per $\Delta\log(\rm SSFR)_{\rm MS}$  bin. The minimum stellar mass  probed thoughtout the different redshift bins is  $\log(M_\star^{\rm lim}/\rm M_\odot)=10.0, 10.2, 10.5, 10.5, \, \rm and \, 10.7$, respectively. \tablefoottext{a}{The highest $\Delta\log(\rm SSFR)_{\rm MS}$  bin is centered at $\Delta\log(\rm SSFR)_{\rm MS}=1.2, 1.05, 1.05, 1.45, 1.45$, respectively.}    }	
		\label{table:3}	
	\end{center} 
\end{table*}

\end{appendix}
\end{document}